\documentclass[prb,aps,10pt,epsfig,longbibliography,twocolumn]{revtex4-2}
\usepackage{graphicx}
\usepackage[unicode=true,colorlinks=true]{hyperref}
\hypersetup{linkcolor=blue,citecolor=blue,urlcolor=blue}
\usepackage{dcolumn}
\usepackage{bm}
\usepackage{ulem}
\usepackage{comment}
\usepackage{amssymb}
\usepackage{hyperref}
\usepackage{mathtools}
\usepackage{comment}
\usepackage{amsmath}  
\usepackage{cleveref}
\usepackage{xcolor}
\usepackage{afterpage}
\usepackage{placeins}
\usepackage[T1]{fontenc}
\renewcommand{\selectlanguage}[1]{}
\DeclareUnicodeCharacter{2212}{\textendash}
\newcommand{\La}{\line (1,0 ){5}}
\newcommand{\Lb}{\line (0,1){5}}
\newcommand{\Ld}{\line (-1,0){5}}
\newcommand{\Le}{\line (0,-1){5}}
\newcommand{\C} {\circle*{2.3}}

\newcommand{\pA}{\put(-2.5,-9)}
\newcommand{\pB}{\put(2.5,-9)}
\newcommand{\pC}{\put(2.5,-3)}
\newcommand{\pZ}{\put(-2.5,-3)}
\newcommand{\rhomb}{
  \pA{\C}\pB{\C}\pZ{\C}\pC{\C}
 }
\newcommand{\rhombH}{
  \begin{picture}(8.1,9)(-3.8,-8.1)
  \linethickness{0.7pt}
    \pB{\Lb}\pZ{\Le}
    \rhomb
  \end{picture}
}
\newcommand{\rhombV}{
  \begin{picture}(8.1,9)(-3.8,-8.1)
  \linethickness{0.7pt}
   \pA{\La}\pC{\Ld}
    \rhomb
  \end{picture}
}
\usepackage{tikz}
\makeatletter
\usepackage{color, colortbl}

\begin{document}
\title{Group Convolutional Neural Network for the Low-Energy Spectrum in the Quantum Dimer Model}
\author{Ojasvi Sharma}
\thanks{These authors contributed equally to this work.}
\author{Sandipan Manna}
\thanks{These authors contributed equally to this work.}
\author{Prashant Shekhar Rao}
\author{G J Sreejith}
\affiliation{Indian Institute of Science Education and Research, Pune, India - 411008}

\begin{abstract}
We obtain the $\rm{p4m}$-symmetric Group Convolutional Neural Network (GCNN) representations of the lowest energy eigenstate of the quantum dimer model on $L{\times} L$ square-lattice in each of the ${(L^2+18L+72)}/{8}$ irreducible representations (irreps) of the lattice space group and use these to investigate the competition between columnar, plaquette and mixed phases. 
The networks are optimized within each irrep by minimizing the energy, which is estimated from samples obtained via an efficient directed loop sampler.
In extensive benchmarks, we show excellent agreement in energy estimates, order parameters and correlation functions with exact diagonalization or quantum Monte Carlo in systems of sizes $8\leq L\leq 32$. 
Analysis of the scaling of the gaps in different representation sectors with systems of sizes up to $L=32$ suggest a $4$-fold degenerate ground state for $V\leq 0.4$ narrowing the regime of possible mixed/plaquette phases to $0.4 < V< 1$. Our results show that GCNN is a powerful tool to investigate ground state phase diagrams. The approach paves the way for even more accurate results by producing highly accurate variational baseline wavefunctions for quantum Monte Carlo approaches.
\end{abstract}

\maketitle
Efficient quantum state representation is an important frontier of research in quantum many-body physics. 
General classes of representations such as tensor networks (TN)~\cite{white_density_1992}, quantum Monte Carlo (QMC) samples~\cite{sandvik_computational_2010}, and shadows~\cite{aaronson_shadow_2020} have distinct regimes of utility and limitations~\cite{loh_sign_1990, henelius_sign_2000, schuch_computational_2007, tagliacozzo_simulation_2009}. 
%
%
Neural quantum states (NQS) employing artificial neural networks (NNs) as variational ansatzes have recently been proposed as scalable representations of many-body states~\cite{carleo_solving_2017,gao_efficient_2017}. 
The universal function approximation property of NNs~\cite{hornik_multilayer_1989} allows NQS, in principle, to represent arbitrary quantum states with sufficient parameterization.
Efficiency, optimizability and expressive power of different NQS architectures are being investigated in several quantum many-body contexts, with Restricted Boltzmann Machines (RBMs) being the focus of early studies due to their simplicity and relation to spin glass models~\cite{carleo_solving_2017, gao_efficient_2017, nomura_restricted_2017, nomura_helping_2021}.
Even simple RBMs can express highly entangled quantum states~\cite{deng_quantum_2017}, offering an alternative to TNs.
Recent works in specific contexts suggest that TN states are a subset of NQS~\cite{zheng_restricted_2019, gao_efficient_2017,sharir_neural_2022,huang_neural_2021,wu_tensor-network_2023} with NQS being capable of efficiently representing certain non-area-law states, thereby going beyond the TN class. 
RBM representations have been utilized to explore ground states of several lattice models, ranging from Ising and Heisenberg systems~\cite{carleo_solving_2017,viteritti_accuracy_2022} to fracton models~\cite{machaczek_neural_2025} and systems possessing non-Abelian and anyonic features~\cite{vieijra_restricted_2020}.
Deep NNs and transformers have been employed to describe fermionic systems and frustrated quantum magnets~\cite{teng_solving_2024, teng_solving_2025, luo_solving_2025, li_deep_2025, chen_neural_2025, rende_transformer_2026, nazaryan_artificial_2026, rende_transformer_2026, qian_describing_2025,fan2026equivariantneuralnetworksforcefield, chen_neural_2025, romero_spectroscopy_2025, viteritti_transformer_2023, raikos2026variationalstudymagnetizationplateaus}.
Motivated by the success in image processing and similarity to coarse-graining~\cite{mehta2014exactmappingvariationalrenormalization}, Convolutional Neural Networks (CNNs) have been used to represent ground states for the $\text{2D\,}J_1 \text{-}J_2$ Heisenberg model~\cite{choo_two-dimensional_2019, liang_solving_2018} up to system size $10 {\times} 10$.
CNNs inherently impose lattice translational symmetries. A generalization incorporating full space group symmetries $-$ Group Convolutional Neural Networks (GCNNs)~\cite{pmlr-v48-cohenc16} were used to study ground and low-lying excited states of $J_1 \text{-}J_2$ model and spin-$\frac{1}{2}$ anti-ferromagnets~\cite{roth2021groupconvolutionalneuralnetworks,roth_high-accuracy_2023,duric_spin-_2025}. 

We use GCNN to study the low-energy states of the Quantum Dimer Model (QDM) on square lattices of linear size up to $L{=}32$ employing two key modifications to the one used in spin systems. 
GCNN for the dimer model here uses a generalization of the GCNN allowing for the local degrees of freedom to transform non-trivially under the lattice space group symmetries. 
Secondly, an efficient Monte Carlo (MC) estimation of the energy for GCNN training is achieved by incorporating a highly efficient infinite temperature directed loop algorithm instead of local updates. 
By optimizing such a GCNN projected into each irreducible representation (irrep) of the lattice space group, we obtain the ground state in each irrep, allowing a careful investigation of the lattice symmetry breaking ground state phase transitions. 
To our knowledge, this is one of the first works~\cite{romero_spectroscopy_2025, PhysRevLett.121.167204} which obtain excited state spectrum using group symmetry structures incorporated into NNs.
In small systems ($L {\leq} 8 $), results obtained from $\mathcal{L} {=}2$ layer GCNN show excellent agreement with Exact Diagonalization (ED), while extensive benchmarking against QMC~\cite{syljuasen_plaquette_2006} show equally good agreement for larger systems up to $L\leq 32$ both in the ground state and excited state sectors.
We find that $\mathcal{L} {=}2$ networks suffice for accurate energies, order parameters, and correlations in these systems. \\
\section{Quantum Dimer Model $(\rm QDM)$}
A paradigmatic model of locally constrained systems is the QDM. 
Originally introduced~\cite{rokhsar_superconductivity_1988} in the context of high-$T_c$ superconductivity as an effective description of resonating valence bond~\cite{moessner_resonating_2001, sutherland_systems_1988} states, QDM is one of the simplest systems exhibiting topological order and fractionalization~\cite{moessner2008quantumdimermodels}. 
Variants of dimer models have been of interest in combinatorics~\cite{allegra_exact_2015, kenyon_double-dimers_2016, jenne_combinatorics_2021, kenyon2009lecturesdimers}, gauge theory and statistical mechanics~\cite{kasteleyn_statistics_1961,moessner_short-ranged_2001, nienhuis_triangular_1984}.
The 2D-QDM Hilbert space is made of classical fully packed hard-core dimer configurations as the basis states, with a Hamiltonian (on a square lattice) given by:
\begin{equation}
H_{\rm QDM}{=}{-t\sum_{\scriptscriptstyle\rm plaq.}} \vert\rhombV\rangle\langle \rhombH\vert{+}\vert\rhombH\rangle\langle \rhombV\vert + {{V}\sum_{\scriptscriptstyle\rm  plaq.}} \vert\rhombV\rangle\langle \rhombV\vert{+}\vert\rhombH\rangle\langle \rhombH\vert
\label{eq:qdm_hamiltonian}
\end{equation}
Here, the first term is the kinetic energy which rotates the flippable plaquettes (plaquettes with two parallel dimers) by $\pi/2$, and the second is a potential energy term that assigns an energy cost $V$ to each flippable plaquette. We set $t=1$ for our calculations.

The QDM phase diagram depends on the lattice geometry and dimensionality~\cite{moessner2008quantumdimermodels,dabholkar_reentrance_2022,yan_sweeping_2019}. 
For $V/t{{>}}1$ (staggered phase), the ground state is highly degenerate and devoid of any flippable plaquette.
As $V/t {\to} -\infty$, the ground state maximizes the flippable plaquettes, yielding a columnar phase with broken rotation and translation symmetry (along one direction).
At the critical RK (Rokhsar-Kivelson) point $V/t{=}1$, the gapless ground state is an equal-amplitude superposition of all dimer configurations. 
The ground state in $V/t\lesssim1$ regime remains unresolved.
Various studies have proposed a range of competing states here ${-}$ columnar, plaquette, and mixed phases~\cite{PhysRevB.54.12938, syljuasen_continuous-time_2005, syljuasen_plaquette_2006, ralko_generic_2008, banerjee_interfaces_2014, banerjee_finite-volume_2016, PhysRevB.98.064302, yan_widely_2021-1, yan_sweeping_2019}. 
This lack of consensus stems primarily from challenging finite size effects and small gaps~\cite{ralko_generic_2008} and motivates the use of complementary methods. 
The current study proposes GCNN ansatz as a possible route to resolving this and other similar questions of the ground state phase diagram. 
We first present careful benchmarking of the ansatz with results from other methods. 
We present the low energy spectrum covering the ground states of all irreps in small systems $L\leq 12$ and the lowest energy irreps relevant to the question of mixed/plaquette/columnar ordering in systems up to $L\leq 32$. 
Analysis of the gap as a function of system size at $V{=}0.4$ suggests collapse of the gap in the irreps consistent with a columnar ordered ground state and a finite gap to plaquette ordering. 
Together with the results for gaps in smaller systems at $V<0.4$ in earlier studies in Ref.~\cite{moessner2008quantumdimermodels}, our results suggest columnar ordering in $V\leq 0.4$ narrowing the possibility of plaquette ordering to $0.4\leq V<1$ which is our primary result. 

\section{Group Convolutional Neural Network} 
The QDM Hamiltonian is invariant under $G\,{=}\,\mathrm{p4m}\,{\cong}\,\mathbb{Z}^2{\rtimes}D_4$ symmetry of the square lattice; every element can be uniquely represented as a composition $t d$ of a translation $t{\in}\mathbb{Z}^2$ (modulo lattice size) and a point group element $d{\in}D_4$ generated by the reflection and $\frac{\pi}{2}$ rotations. 
Each dimer configuration can be uniquely written as a map $\sigma\,{:}\,\mathbb{L}{\rightarrow}O$, from the lattice $\mathbb{L}$ to the set of possible dimer orientations $O\,{=}\, \{ \pm\hat{x},\pm\hat{y} \}$ on each site $\vec r\,{\in}\,\mathbb{L}$. 
The symmetry element $g\,{=}\,td$ acts on a configuration $\sigma$ as $(g\sigma)_{\vec r}\,{=}\,d(\sigma_{(g^{-1}\vec{r})})$ where $g$-action on $\vec r$ and $d$-action on the local state are the natural actions of $\rm p4m$ and $D_4$ on $\mathbb{L}$ and $O$ respectively. 
Ground state of the QDM transforms under the trivial representation of the ${\rm p4m}$. 
Imposing this constraint explicitly through the GCNNs improves the ansatz accuracy~\cite{dascoli2020findingneedlehaystackconvolutions,pmlr-v48-cohenc16, roth2021groupconvolutionalneuralnetworks,PhysRevB.109.054410, roth_high-accuracy_2023,PhysRevLett.127.276402}. 
Note that unlike the GCNNs studied earlier where configurations transformed as a scalar, the local states here transform non-trivially under $G$.

First layer of the GCNN maps the input configuration $\sigma$ to a feature map $\text{f}^1$ over $G$:
\begin{equation}
\label{eq:embedding_layer}
    \boldsymbol{\text{f}}_g^{1,\alpha} = \Gamma \left[ \sum_{\vec{r}\in \mathbb{L}} (g^{-1}\sigma)_{\vec{r}} K^{\alpha}_{\vec{r}} + b^{1,\alpha}\right],\;g\in {G}
\end{equation}
where ${K:\mathbb{L} \rightarrow \mathbb{C} }$ is the embedding kernel, ${\boldsymbol{\text{f}}^1 : G \rightarrow \mathbb{C}}$ is the feature map obtained from the first layer, $b^1$ is the bias, and $\Gamma$ is a non-linear activation function chosen as the complex version of the Scaled Exponential Linear Unit (SELU)~\cite{klambauer_self-normalizing_2017}  - for $x \in \mathbb{C}$, $\Gamma(x) = \text{SELU}(\Re(x)) + i \text{SELU}(\Im(x)) $. 
The superscript ${\alpha=1,\dots c_1}$ is the channel index. For brevity, hereafter we suppress the channel indices.
The embedding layer satisfies the following equivariance for any $u,g\in G$ and any configuration $\sigma$:
\begin{equation}
\text{f}_g(u\sigma)=\text{f}_{u^{-1}g}(\sigma).\label{eq:embeddingEquivariance}
\end{equation}
\begin{figure}
    \includegraphics[width=0.9\linewidth]{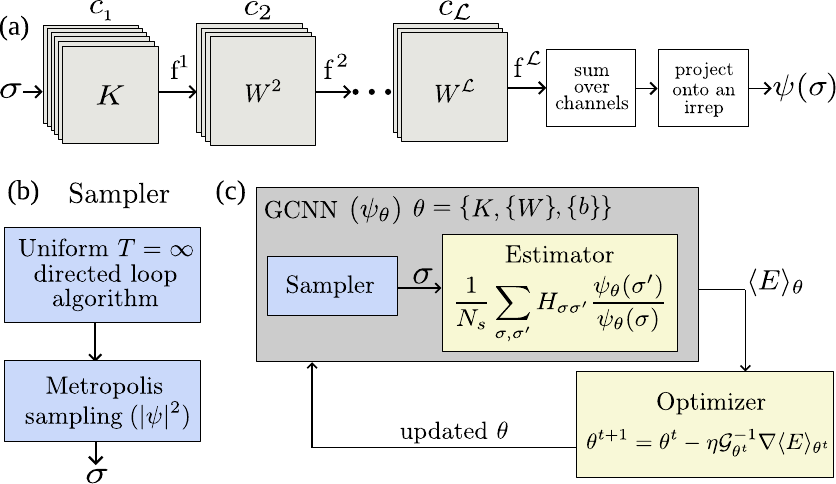}
    \caption{(a) GCNN architecture with $\mathcal{L}$ layers.\\(b) Working of the sampler. (c) Schematic of the algorithm.
    \label{fig:schematics}}
\end{figure}
The output of the embedding layer is further passed to $(\mathcal{L}{-}1)$ group-convolutional (GC) layers ($\mathcal{L}$ is the number of layers) which take one feature map over $G$ into another:
\begin{equation}
\label{eq:gconv_layer}
    \boldsymbol{\text{f}}_g^{\,i+1} = {\rm GC}[ \boldsymbol{\text{f}}^{\,i}]_g\coloneq\Gamma \left[ \sum_{h \in G} W_{h^{-1}g}^{i+1} \boldsymbol{\text{f}}_h^{\,i} + b^{i+1}\right].
\end{equation}
Here $W,b$ are the GC kernels and biases. The GCs (\ref{eq:gconv_layer}) satisfy the following equivariance $\forall u\in G$:
\begin{align}
   {\rm GC}[ \boldsymbol{\text{f}}^{\,i}\circ u]_g=\Gamma \left[ \sum_{h \in {\rm G}} W^{i+1}_{h^{-1}g} \boldsymbol{\text{f}}^{\,i}_{uh}{+}b^{i+1} \right]=\boldsymbol{\text{f}}^{\,i+1}_{ug} \label{eq:gConvEquivariance}
\end{align}
The features at the $\mathcal{L}^{\text{th}}$ layer are added and projected on an irreducible representation (irrep) of $G$ as:
\begin{equation}
\label{eq:project_on_irreps}
    \psi(\sigma) = \sum_{g\in G} \chi_{g}^* \sum_{k=1}^{c_{\mathcal{L}}}  \exp\, \boldsymbol{\text{f}}^{\,\mathcal{L},k}_{g} 
\end{equation}
where $\chi_{g}$ are the characters of the desired irrep $\rho$, and $c_{\mathcal{L}}$ is the number of channels in the $\mathcal{L}^{\text{th}}$ layer.
Equivariances (Eq.~\ref{eq:embeddingEquivariance}, \ref{eq:gConvEquivariance}) and Eq.~\ref{eq:project_on_irreps} imply $\psi(\sigma)$ transforms in the selected irrep $\rho$.
An $\mathcal{L}$-layer GCNN has $c_1 L^2 + \sum_{i=1}^{\mathcal{L}-1}c_i c_{i+1} |G| + \sum_{i=1}^{\mathcal{L}}c_i$ complex parameters where 
$|G|$ is the order of the group ($8L^2$ for $\rm{p4m}$ group), and $c_i$ is the number of channels in the $i^{\rm th}$layer. 
We use  $\mathcal{L}{=}1,2,3$ layers with $ \left( 2\mathcal{L},\ 2(\mathcal{L}{-}1), \dots,\ 2 \right)$ channels in the layers, following the choice in Ref.~\onlinecite{choo_two-dimensional_2019}. 
Thus, networks with $\mathcal{L}{=}1, 2$ and $3$ have $(2), (4,2)$ and $(6,4,2)$ channels in each layer, respectively. At $L{=}8$, there are 130, 4358, 16780 complex parameters respectively. We note that the sign structure of the Hamiltonian forces the global ground state in the fully symmetric irrep to be a non-negative  vector allowing use of real network parameters, however ground states in general irreps require complex parameters as the Hamiltonian projected into a general irrep is not of Frobenius type.

Figure~\ref{fig:schematics}(a) shows the $\mathcal{L}$-layer GCNN which produces the $\rm{p4m}$-invariant function $\psi_\theta(\sigma)$ parametrized by the weights and biases $\theta$ and 
Fig.~\ref{fig:schematics}(c) presents the optimization scheme~\cite{carleo_netket_2019, vicentini_netket_2022} (see Section 4.1 of ~\cite{vicentini_netket_2022} for details). 
Given $\sigma$, the $T{=}\infty$ directed loop sampler (Fig.~\ref{fig:schematics}(b)) proposes $\sigma'$ which is accepted with Metropolis probability, yielding $|\psi_\theta|^2$-distributed samples upon iteration. 
The energy expectation $\langle E \rangle_{\theta}$, estimated as the average of $E_{\text{loc}} (\sigma) {=}\sum_{\sigma'} H_{\sigma \sigma'}{\psi_{\theta} (\sigma') }/{ \psi_{\theta} (\sigma)}$ over the MC samples, is used as the cost function to update the parameters using stochastic reconfiguration. 
In the following discussion, we present an extensive comparison of the GCNN with ED and QMC results. 
All networks are optimized by minimizing the estimated energy using $2^{12}$ MC samples (up to $2^{14}$ MC samples for $L>16$). 
The expectation values of observables are estimated with $2^{20}$ MC samples.

\begin{figure}[h]
	\includegraphics[width=0.9\columnwidth]{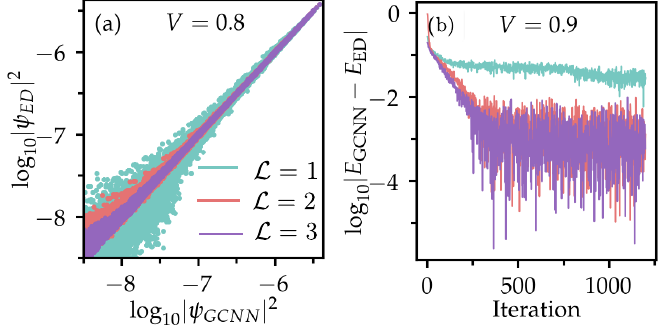}
	\caption{Benchmarking GCNN results against ED for $L{=}8$. (a) Comparison of $\psi(\sigma)$ from GCNN and ED at $V{=}0.8$. (b) Convergence of energy with iterations : $\log_{10}\Delta E$ vs. iterations at $V{=}0.9$. 
		\label{fig:ed_l8_maintext}}
\end{figure}
\FloatBarrier

\section{Comparison with ED and QMC}
We benchmark our $\rm{GCNN}$ results against ED and QMC across a range of system sizes to ensure that the choices of the network architecture, Monte Carlo estimators, thermalization steps, sweep sizes for sampling, uniform samplers over basis states etc do not result in systematic biases. The highly efficient directed loop sampler ensure that the MC samplers are ergodic even for the largest systems we investigate. 

\subsection{Comparison with ED} 
\begin{figure*}[ht]
	\centering
	\includegraphics[width=0.8\textwidth]{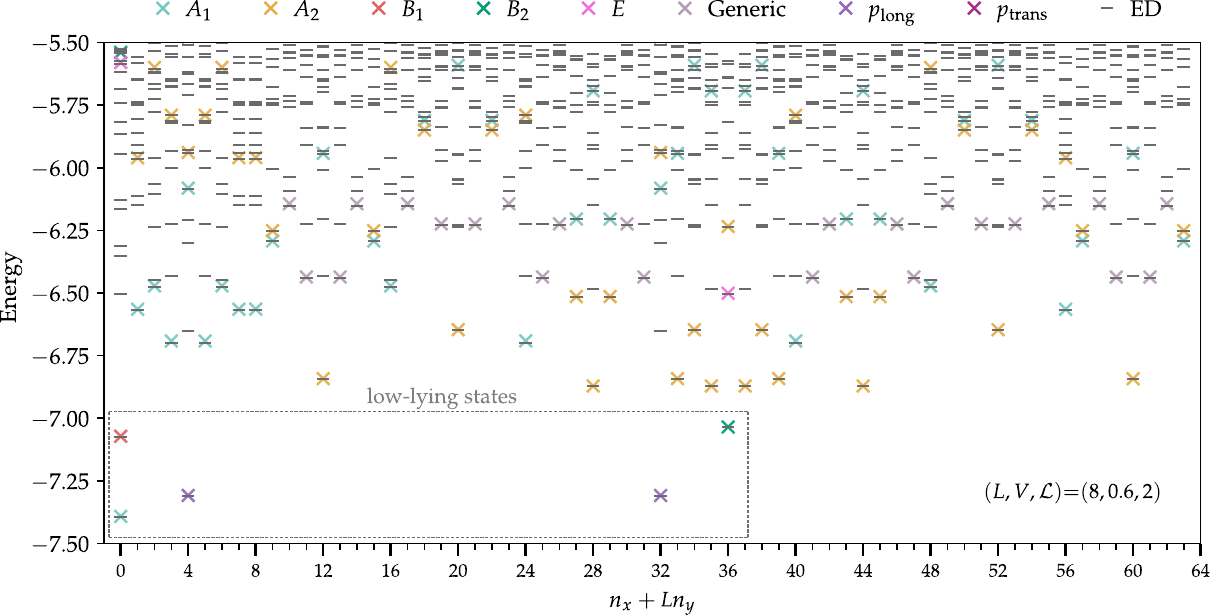}
	\caption{Momentum- and symmetry-resolved energy spectrum at $V{=}0.6$ obtained with $\mathcal{L}{=}2$ GCNN across the $\rm p4m$ symmetry group for $L{=}8$: Energy eigenvalues are shown as a function of the crystal momentum $\vec{k} =(k_x, k_y)$, where $k_{x(y)} = 2 \pi n_{x(y)}/L$. Black dashed markers denote ED eigenvalues for each momentum point across all winding and symmetery sectors. Colored crosses indicate energies obtained from GCNN, with colors corresponding to different irreps of the lattice symmetry group. For every momentum point, the GCNN correctly captures the ground-state energy in each symmetry sector. The dotted box highlights the global ground state (in $A_1$ irrep and $(0,0)$ momentum sector) and low-lying states below the gap.}
	\label{fig:momentum_spectrum_L8}
\end{figure*}

At $L{=}8$, we compare the global ground state wavefunction amplitudes as well as every irrep-specific ground state energy with ED (See Fig.~\ref{fig:momentum_spectrum_L8}) and find excellent agreement throughout. In Fig.~\ref{fig:ed_l8_maintext}, we compare the GCNN ground state with that from ED at $L{=}8$ . The full Hilbert space dimension is $311853312$ but when calculating the ground state in the trivial irrep, GCNN performs optimization within the 628931 dimensional ${\rm p4m}$-invariant subspace. In Fig.~\ref{fig:ed_l8_maintext}(a) we compare $628931$ values of $|\psi(\sigma)|^2$ obtained from GCNN and ED for $V{=}0.8$. Small-magnitude coefficients improve with $\mathcal{L}$ resulting in better fidelity (See Appendix Fig.~\ref{fig:L8_ener_and_error_coeffs} for other $V$ values). Energy obtained from optimized GCNN, $E_{\rm GCNN}$ converges exponentially to the corresponding energy values obtained from $E_{\rm ED}$ with iterations (Fig.~\ref{fig:ed_l8_maintext}(b)) until saturation likely due to limitations of MC sampling.

\begin{figure}
	\includegraphics[width=0.95\columnwidth]{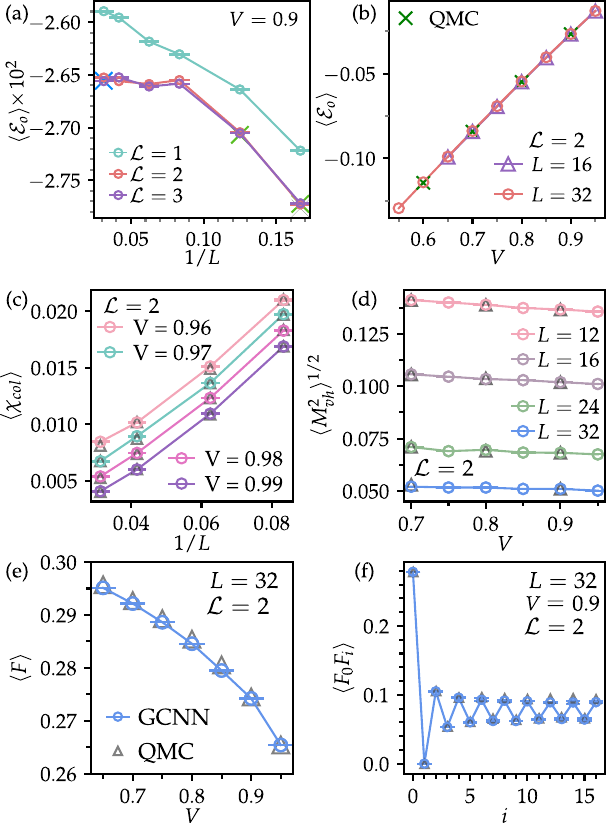}
	\caption{Benchmarking GCNN results against QMC~\cite{syljuasen_plaquette_2006} for  $12 \leq L\leq 32$. (a) Energy density $\langle \mathcal{E}_o \rangle$ vs $1/L$ with mean and errorbars estimated from final 100 optimization iterations after convergence. Blue and green crosses mark QMC and ED results, respectively.
		(b) $\langle \mathcal{E}_o \rangle$ as a function of $V$.
		(c) $\langle \chi_{\rm col}\rangle$ vs $1/L$ for different $V$s. 
		(d) Nematic order parameter vs $V$ for $12 \leq L\leq 32$. 
		(e) Flip operator per plaquette as function of $V$ for $L{=}32$. 
		(f) Two-point correlation of flip operator as a function of diagonal separation $\vec{r_i} {=} (i, i)$ at $(L,V){=}(32, 0.9)$. (For (c)-(f): circles and triangles represent $\rm GCNN$ and $\rm QMC$ results respectively.) The errorbars are smaller than the marker size.
		\label{fig:merged_plot}}
\end{figure}

\subsection{Comparison with QMC}
\paragraph*{Energies: }
For larger system sizes, we benchmark our results against QMC results from Ref.~\cite{syljuasen_plaquette_2006,ralko_generic_2008} in systems up to $L=32$. 
In Fig.~\ref{fig:merged_plot}(a), we compare the ground-state energy density $\langle \mathcal{E}_o \rangle$ estimated from GCNNs of varying depths with QMC estimates~\cite{syljuasen_plaquette_2006} at a fixed $V$. 
While $\mathcal{L} {=} 1$ network performs poorly, $\mathcal{L} {\geq} 2$ networks agree with QMC across system sizes. 
Figure~\ref{fig:merged_plot}(b) shows $\mathcal{L}{=}2$ $\text{GCNN}$ estimates of $\langle \mathcal{E}_o \rangle$ as a function of $V$ at $L{=}16,32$, demonstrating its sufficiency across $V$. 
In Fig.~\ref{fig:spectrum_maintext}(c), we show a comparison of $\mathcal{L}{=}2$ GCNN estimates of the low lying energies with the corresponding QMC estimates~\cite{ralko_generic_2008} at $V{=}0.4$ across $L$ again indicating excellent agreement even for excitations.

\paragraph*{Columnar and Nematic order parameters (OPs):} 
To gauge the fidelity of the GCNN results beyond energy estimates, we compare order parameters with those estimated from QMC~\cite{syljuasen_plaquette_2006}. 
We first consider the columnar OP 
$\vec{C}{=}(C_x , C_y)$,
where $2L^2C_{\mu}{=}\sum_{\vec{r}}(-1)^{r_{\mu}}\left[n_{\mu}(\vec{r}){-}n_{-\mu}(\vec{r})\right]$ , $\mu {\in} \{x,y\}$, where $n_{\mu}(\vec{r}){=}1$ iff a dimer exists at $\vec{r}$ pointing in the $\mu$ direction, and zero otherwise. We show $\langle \chi_{{col}} \rangle {=} \langle |\vec{C}|^2 \rangle$ at $V$ close to the RK point for different $L$ in Fig.~\ref{fig:merged_plot}(c). 
We also consider the nematic OP defined as $M_{vh} {=} \frac{2}{L^2} (N_v - N_h)$ where $N_v$ ($N_h$) is the number of vertical (horizontal) dimers. We present $\langle M_{vh}^2 \rangle^{1/2}$ for various $V$ close to the RK point across system sizes in Fig.~\ref{fig:merged_plot}(d). 
Estimates of both OPs agree remarkably well with the corresponding QMC results even near the critical point. These OPs, diagonal in the dimer basis, are evaluated on $N_s \sim 2^{20}$ MC samples from the optimized GCNN state.

\paragraph*{Plaquette flip operator and correlation function:} All observables compared so far, except energy, are diagonal in the dimer basis. To benchmark off-diagonal observables, we consider the flip operator $F{=}\frac{1}{L^2}{\sum_{\scriptscriptstyle\rm plaq.}} \vert\rhombV\rangle\langle \rhombH\vert{+}\vert\rhombH\rangle\langle \rhombV\vert$. While its expectation value can be estimated from $\mathcal{E}_o$ and a diagonal observable as $\langle F \rangle {=}-\langle \mathcal{E}_{o}\rangle{+}V \frac{\langle N_f \rangle}{L^2}$, 
where $\langle N_f \rangle$ is the number of flippable plaquettes (Fig.~\ref{fig:merged_plot}(e)), the two-point correlation function $\langle F_0 F_i \rangle$ has off-diagonal components independent of $\langle \mathcal{E}_{o}\rangle$.
The result, shown in Fig.~\ref{fig:merged_plot}(f) for $(L, V){=}(32, 0.9)$, exhibits oscillations with slowly decaying envelope. All the results we benchmarked are in excellent agreement with QMC, indicating the sufficiency across $V,L$ of a $\mathcal{L}{=}2$ GCNN with $(4,2)$ channels.

\section{Space group resolved energy spectrum}
The $\rm p4m$ group restricted to the finite $L {\times} L$ square lattice has $(L^2+18L+72)/8$ distinct irreps, each labeled by a Brillouin zone momentum (modulo $D_4$) $k{=}(k_x,k_y)$ and a label characterizing a representation of the little group at $k$ ( stabilizer for $D_4$ action on $k$). 
The group characters $\chi_g^\rho ~ \forall g \in G$ in each irrep $\rho$ can be calculated by techniques of induced representations (See Appendix~\ref{app:p4m_irreps}) and can be used to construct GCNN state in a chosen irrep. In Fig.~\ref{fig:momentum_spectrum_L8}, we present the momentum- and symmetry- resolved energy spectrum obtained from $\mathcal{L}{=}2$ $\rm GCNN$ for $L{=}8, V{=}0.6$ and compare it with the spectrum obtained from ED across different winding and momentum sectors. 
The eigenstates are characterized by distinct quantum numbers, defined by a momentum and a little group irrep. The global ground state lies in the $(0,0)$ momentum sector and the $A_1$ irrep. We find that the $\rm GCNN$ correctly captures the set of ground states (cross markers with colors corresponding to the different irreps) in each irrep and they match excellently with $\rm ED$ results (black dashed markers). 

\begin{figure}{
    \includegraphics[width=\columnwidth]{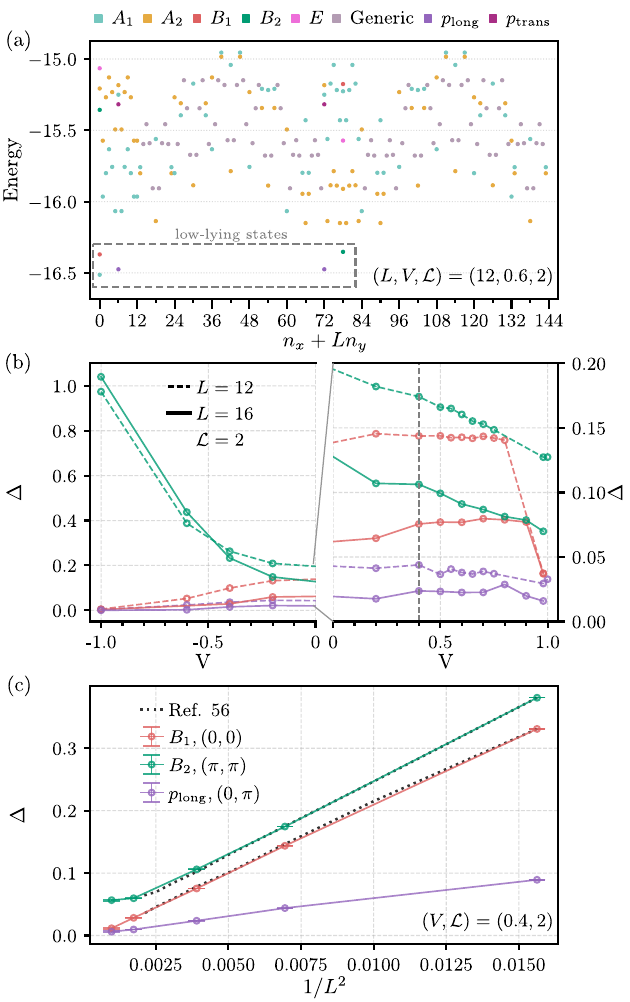}
    \caption{(a) Energy vs momentum (labeled by $n_\mu{=}k_\mu L/2\pi$) at $V{=}0.6$ obtained with GCNN for $L{=}12$. The low-lying states are highlighted inside the dashed box. (b) Excitation gaps ($\Delta$) for low-lying states as a function of $V$ for $L{=}12$ (dashed lines) and $16$ (solid lines) (c) $\Delta$ for low-lying states vs $L$ at $V{=}0.4$. The dotted line shows results from Ref.~\cite{ralko_generic_2008}, demonstrating excellent agreement in $L{\leq}22$ where data is available.}
    \label{fig:spectrum_maintext}}
\end{figure}

Having validated the GCNN estimate of excitation energies for $L=8$ against exact spectrum from ED, we next study larger systems. Figure~\ref{fig:spectrum_maintext}(a) shows the lowest energies, at $L{=}12$, calculated by minimizing the energy of the GCNN projected onto each irrep. Though this does not capture all low energy eigenvalues i.e., excited states in each sector are not produced, it contains the necessary information to detect space group symmetry breaking. 

\subsection{Irreps in the ground state space associated with different broken symmetry phases}
The columnar order is associated with a 4-fold symmetry broken phase which breaks the discrete translation symmetry in one of the directions ($\mathbb{Z}$ to $2\mathbb{Z}$) and the rotational symmetry ($C_4$ symmetry to $C_2$). The space of the $4$ columnar ordered states decomposes into $3$ irreducible representations of the symmetry group namely
\begin{align}
A_1,k{=}(0,0)&: \sigma_{h,0}+\sigma_{v,0}+\sigma_{h,1}+\sigma_{v,1}\nonumber\\
B_1, k{=}(0,0)&: \sigma_{h,0}+\sigma_{h,1}-\sigma_{v,0}-\sigma_{v,1}\\
p_{\rm long},k{=}(0,\pi){\rm mod} D_4&: {\rm Span}\{\sigma_{h,0}-\sigma_{h,1},\sigma_{v,0}-\sigma_{v,1}\}\nonumber
\end{align}
where $\sigma_{h,0}$($\sigma_{v,0}$) is the columnar ordered state with dimers oriented in the $x$($y$) direction and $\sigma_{\cdot,1}$ is obtained from $\sigma_{\cdot,0}$ by a unit translation perpendicular to the dimers (Appendix~\ref{app:columnar_plaquette_state_illustration}). $A_1$ is the trivial irrep and $B_1$ transforms non-trivially under $C_4$. $p_{\rm long}$ is a 2D irrep which transforms non-trivially under $C_4$ and unit translations. Columnar ordering is thus associated with 4 degenerate states at momenta $(0,0), (0,\pi), (\pi,0)$. Plaquette states breaks the translation symmetry to $2\mathbb{Z}{\times}2\mathbb{Z}$, the space of the plaquette states decomposes into three irreps 
\begin{align}
A_1,k{=}(0,0)&: \psi_{0,0}+\psi_{1,0}+\psi_{0,1}+\psi_{1,1}\nonumber\\
B_2, k{=}(\pi,\pi)&: \psi_{0,0}-\psi_{1,0}-\psi_{0,1}+\psi_{1,1}\\
p_{\rm long},k{=}(0,\pi){\rm mod} D_4&: {\rm Span}\{\psi_{0,0}-\psi_{1,1},\psi_{1,0}-\psi_{0,1}\}.\nonumber
\end{align}
Ground state degeneracy associated with the plaquette phase results in low energy states at momenta $(0,0),(0,\pi), (\pi,0)$ and $(\pi,\pi)$. 

Figure~\ref{fig:spectrum_maintext}(a) shows that the low energy states below the gap have irrep labels consistent with both columnar and plaquette phases discussed above - pointing to the close competition between the two phases.
Across parameter regimes and system sizes, we find that the global ground state at finite system sizes is always the $A_1$ state at $(0,0)$. To determine the ground state phase, we investigate the gap between the $A_1$ ground state and the ground states in the remaining four sectors.

Figure~\ref{fig:spectrum_maintext}(b) shows the energies of the low-energy $B_1$, $B_2$ and $p_{\rm long}$ states  relative to the $A_1$ state, denoted by $\Delta_{B_1}$, $\Delta_{B_2}$ and $\Delta_{p}$, respectively, as a function of $V$ for systems $L{=}12$ and $16$. 
At large negative $V$, consistent with the expected columnar phase, $\Delta_{B_1}$ and $\Delta_{p}$ approach zero within numerical accuracy, indicating degeneracy with $A_1$ state, while $\Delta_{B_2}$ is much higher. 
As $V$ increases, $\Delta_{B_2}$ decreases, indicating competition between different phases in this regime. 
To understand the fate of the columnar phase as $V$ increases, we focus on $V{=}0.4$ and analyze the scaling of $\Delta$ with $L$ (Fig.~\ref{fig:spectrum_maintext}(c)). For $L{=}32$, the GCNN was optimized over $1200$ iterations using $8192$ MC samples, followed by an additional $200$ iterations with the sample size increased to $16384$. We find that $\Delta$ in different symmetry sectors are in good agreement with those reported in Ref.~\cite{ralko_generic_2008}, where the calculations were performed up to $L{=}22$. In particular, $\Delta_{B_1}$ and $\Delta_{p}$ decrease rapidly with increasing $L$ ruling out a pure plaquette phase. $\Delta_{B_2}$ also decreases with $L$  but shows a saturation albeit at a small value for $L{\geq} 20$ suggesting that as $V$ is increased, the columnar phase persists till $V{=}0.4$ at least.

\section{Conclusions}
We have demonstrated the utility of the GCNN ansatz for capturing low-energy spectra of systems with a nontrivial space group symmetries. We have provided extensive benchmarks with exact diagonalization, as well as comparisons with existing QMC results, to validate the approach as well as to establish the adequacy of the hyper-parameters and sampling algorithms used in the our calculations. In addition to demonstrating high fidelity with ED results ($L=8$), we have presented comparison (with available QMC results) of energies, OP expectation values and two point correlations in the global ground state in systems as large as $L=32$. We compared the irrep resolved energies against ED ($L=8$) and QMC (till $L=22$). These tests provide systematic validation of the methods before their application in larger systems. 

We showed that the access to irrep resolved energies can be used to detect the phase diagram with broken space group symmetry phases. We have used the methods developed and validated in this work to show that the ground state of the quantum dimer model has columnar order at least in the $V\leq 0.4$ regime narrowing down the region of mixed phase to $1>V>0.4$. The excitation gaps to be resolved decrease with increasing $V$ making the approach more expensive to use for larger $V$. Estimating the thermodynamic gaps here requires larger system sizes and improved numerical precision. The calculations presented here were obtained using networks trained using A30 and V100 GPUs with up to 24GB memory. 
With the availability of more powerful computing systems, we expect to be able to produce accurate energy estimates in larger systems as well with little modification of the current techniques. 

Finally, we note here that projection Monte Carlo methods built on GCNN trial wavefunctions provide a promising offshoot of the techniques developed here. They will allow us to obtain more accurate estimates of excitation gaps even with limited computing resources for training. Such hybrid methods can also help reduce any variational bias due to finite network complexity. We are currently developing techniques in this direction, and results will be reported in a future publication. The network architecture and the directed loop sampler that supplement the NetKet codes as well as the weights for the trained networks can be found in \url{https://github.com/sgj138/GCNN_Dimer.git}.
\begin{acknowledgments}
We thank O Syljuasen for sharing numerical results from calculations in previous publications. GJS thanks D Kalra, S Powell, F Alet and K Damle for valuable discussions in previous projects.
We acknowledge the use of NetKet~\cite{vicentini_netket_2022, carleo_netket_2019}  in performing the numerical calculations presented in this work.
We thank National Supercomputing Mission for providing computing resources of ``PARAM Brahma'' at IISER Pune, implemented by C-DAC and supported by the Ministry of Electronics and Information Technology (MeitY) and Department of Science and Technology (DST), Government of India. 
GJS thanks the I-HUB Quantum Technology Foundation for financial support. 
OS acknowledges financial support from the Council of Scientific and Industrial Research (CSIR), India, in the form of Junior Research Fellowship 
09/0936(19331)/2024-EMR-I.
\end{acknowledgments}

\typeout{}
\bibliography{qdm_recent}

@article{PhysRevB.109.054410,
  title = {Noncoplanar and chiral spin states on the way towards N\'eel ordering in fullerene Heisenberg models},
  author = {Szab\'o, Attila and Capponi, Sylvain and Alet, Fabien},
  journal = {Phys. Rev. B},
  volume = {109},
  issue = {5},
  pages = {054410},
  numpages = {15},
  year = {2024},
  month = {Feb},
  publisher = {American Physical Society},
  doi = {10.1103/PhysRevB.109.054410},
  url = {https://link.aps.org/doi/10.1103/PhysRevB.109.054410}
}

@article{PhysRevLett.127.276402,
  title = {Gauge Equivariant Neural Networks for Quantum Lattice Gauge Theories},
  author = {Luo, Di and Carleo, Giuseppe and Clark, Bryan K. and Stokes, James},
  journal = {Phys. Rev. Lett.},
  volume = {127},
  issue = {27},
  pages = {276402},
  numpages = {6},
  year = {2021},
  month = {Dec},
  publisher = {American Physical Society},
  doi = {10.1103/PhysRevLett.127.276402},
  url = {https://link.aps.org/doi/10.1103/PhysRevLett.127.276402}
}

@article{PhysRevB.54.12938,
  title = {Columnar dimer and plaquette resonating-valence-bond orders in the quantum dimer model},
  author = {Leung, P. W. and Chiu, K. C. and Runge, Karl J.},
  journal = {Phys. Rev. B},
  volume = {54},
  issue = {18},
  pages = {12938--12945},
  numpages = {0},
  year = {1996},
  month = {Nov},
  publisher = {American Physical Society},
  doi = {10.1103/PhysRevB.54.12938},
  url = {https://link.aps.org/doi/10.1103/PhysRevB.54.12938}
}

@article{PhysRevB.98.064302,
  title = {Phases of quantum dimers from ensembles of classical stochastic trajectories},
  author = {Oakes, Tom and Powell, Stephen and Castelnovo, Claudio and Lamacraft, Austen and Garrahan, Juan P.},
  journal = {Phys. Rev. B},
  volume = {98},
  issue = {6},
  pages = {064302},
  numpages = {13},
  year = {2018},
  month = {Aug},
  publisher = {American Physical Society},
  doi = {10.1103/PhysRevB.98.064302},
  url = {https://link.aps.org/doi/10.1103/PhysRevB.98.064302}
}

@article{dabholkar_reentrance_2022,
  title    = {Reentrance effect in the high-temperature critical phase of the quantum dimer model on the square lattice},
  volume   = {106},
  issn     = {2469-9950, 2469-9969},
  url      = {https://link.aps.org/doi/10.1103/PhysRevB.106.205121},
  doi      = {10.1103/PhysRevB.106.205121},
  language = {en},
  number   = {20},
  urldate  = {2025-04-18},
  journal  = {Physical Review B},
  author   = {Dabholkar, Bhupen and Sreejith, G. J. and Alet, Fabien},
  month    = nov,
  year     = {2022},
  pages    = {205121}
}

@misc{roth2021groupconvolutionalneuralnetworks,
      title={Group Convolutional Neural Networks Improve Quantum State Accuracy}, 
      author={Christopher Roth and Allan H. MacDonald},
      year={2021},
      eprint={2104.05085},
      archivePrefix={arXiv},
      primaryClass={quant-ph},
      url={https://arxiv.org/abs/2104.05085}, 
}

@inproceedings{pmlr-v48-cohenc16,
  title={Group equivariant convolutional networks},
  author={Cohen, Taco and Welling, Max},
  booktitle={International conference on machine learning},
  pages={2990--2999},
  year={2016},
  url={https://proceedings.mlr.press/v48/cohenc16.html},
  organization={PMLR}
}

@misc{dascoli2020findingneedlehaystackconvolutions,
      title={Finding the Needle in the Haystack with Convolutions: on the benefits of architectural bias}, 
      author={Stéphane d'Ascoli and Levent Sagun and Joan Bruna and Giulio Biroli},
      year={2020},
      eprint={1906.06766},
      archivePrefix={arXiv},
      primaryClass={cs.LG},
      url={https://arxiv.org/abs/1906.06766}, 
}

@misc{moessner2008quantumdimermodels,
      title={Quantum dimer models}, 
      author={R. Moessner and K. S. Raman},
      year={2008},
      eprint={0809.3051},
      archivePrefix={arXiv},
      primaryClass={cond-mat.str-el},
      url={https://arxiv.org/abs/0809.3051}, 
}

@article{syljuasen_continuous-time_2005,
  title     = {Continuous-time diffusion {Monte} {Carlo} method applied to the quantum dimer model},
  volume    = {71},
  copyright = {http://link.aps.org/licenses/aps-default-license},
  issn      = {1098-0121, 1550-235X},
  url       = {https://link.aps.org/doi/10.1103/PhysRevB.71.020401},
  doi       = {10.1103/PhysRevB.71.020401},
  language  = {en},
  number    = {2},
  urldate   = {2025-04-18},
  journal   = {Physical Review B},
  author    = {Syljuåsen, Olav F.},
  month     = jan,
  year      = {2005},
  pages     = {020401}
}

@article{syljuasen_plaquette_2006,
  title      = {Plaquette phase of the square-lattice quantum dimer model: {Quantum} {Monte} {Carlo} calculations},
  volume     = {73},
  copyright  = {http://link.aps.org/licenses/aps-default-license},
  issn       = {1098-0121, 1550-235X},
  shorttitle = {Plaquette phase of the square-lattice quantum dimer model},
  url        = {https://link.aps.org/doi/10.1103/PhysRevB.73.245105},
  doi        = {10.1103/PhysRevB.73.245105},
  language   = {en},
  number     = {24},
  urldate    = {2025-04-18},
  journal    = {Physical Review B},
  author     = {Syljuåsen, Olav F.},
  month      = jun,
  year       = {2006},
  pages      = {245105}
}

@article{ralko_generic_2008,
  title     = {Generic {Mixed} {Columnar}-{Plaquette} {Phases} in {Rokhsar}-{Kivelson} {Models}},
  volume    = {100},
  copyright = {http://link.aps.org/licenses/aps-default-license},
  issn      = {0031-9007, 1079-7114},
  url       = {https://link.aps.org/doi/10.1103/PhysRevLett.100.037201},
  doi       = {10.1103/PhysRevLett.100.037201},
  language  = {en},
  number    = {3},
  urldate   = {2025-04-18},
  journal   = {Physical Review Letters},
  author    = {Ralko, A. and Poilblanc, D. and Moessner, R.},
  month     = jan,
  year      = {2008},
  pages     = {037201}
}

@article{banerjee_interfaces_2014,
  title     = {Interfaces, strings, and a soft mode in the square lattice quantum dimer model},
  volume    = {90},
  copyright = {http://link.aps.org/licenses/aps-default-license},
  issn      = {1098-0121, 1550-235X},
  url       = {https://link.aps.org/doi/10.1103/PhysRevB.90.245143},
  doi       = {10.1103/PhysRevB.90.245143},
  language  = {en},
  number    = {24},
  urldate   = {2025-04-18},
  journal   = {Physical Review B},
  author    = {Banerjee, D. and Bögli, M. and Hofmann, C. P. and Jiang, F.-J. and Widmer, P. and Wiese, U.-J.},
  month     = dec,
  year      = {2014},
  pages     = {245143}
}

@article{banerjee_finite-volume_2016,
  title     = {Finite-volume energy spectrum, fractionalized strings, and low-energy effective field theory for the quantum dimer model on the square lattice},
  volume    = {94},
  copyright = {http://link.aps.org/licenses/aps-default-license},
  issn      = {2469-9950, 2469-9969},
  url       = {https://link.aps.org/doi/10.1103/PhysRevB.94.115120},
  doi       = {10.1103/PhysRevB.94.115120},
  language  = {en},
  number    = {11},
  urldate   = {2025-04-18},
  journal   = {Physical Review B},
  author    = {Banerjee, D. and Bögli, M. and Hofmann, C. P. and Jiang, F.-J. and Widmer, P. and Wiese, U.-J.},
  month     = sep,
  year      = {2016},
  pages     = {115120}
}

@article{yan_widely_2021-1,
  title    = {Widely existing mixed phase structure of the quantum dimer model on a square lattice},
  volume   = {103},
  issn     = {2469-9950, 2469-9969},
  url      = {https://link.aps.org/doi/10.1103/PhysRevB.103.094421},
  doi      = {10.1103/PhysRevB.103.094421},
  language = {en},
  number   = {9},
  urldate  = {2025-04-18},
  journal  = {Physical Review B},
  author   = {Yan, Zheng and Zhou, Zheng and Syljuåsen, Olav F. and Zhang, Junhao and Yuan, Tianzhong and Lou, Jie and Chen, Yan},
  month    = mar,
  year     = {2021},
  pages    = {094421}
}

@article{rokhsar_superconductivity_1988,
  title     = {Superconductivity and the {Quantum} {Hard}-{Core} {Dimer} {Gas}},
  volume    = {61},
  copyright = {http://link.aps.org/licenses/aps-default-license},
  issn      = {0031-9007},
  url       = {https://link.aps.org/doi/10.1103/PhysRevLett.61.2376},
  doi       = {10.1103/PhysRevLett.61.2376},
  language  = {en},
  number    = {20},
  urldate   = {2025-04-18},
  journal   = {Physical Review Letters},
  author    = {Rokhsar, Daniel S. and Kivelson, Steven A.},
  month     = nov,
  year      = {1988},
  pages     = {2376--2379}
}

@article{carleo_solving_2017,
  title     = {Solving the quantum many-body problem with artificial neural networks},
  volume    = {355},
  url       = {https://www.science.org/doi/10.1126/science.aag2302},
  doi       = {10.1126/science.aag2302},
  number    = {6325},
  urldate   = {2025-04-20},
  journal   = {Science},
  publisher = {American Association for the Advancement of Science},
  author    = {Carleo, Giuseppe and Troyer, Matthias},
  month     = feb,
  year      = {2017},
  pages     = {602--606}
}

@article{choo_two-dimensional_2019,
  title    = {Two-dimensional frustrated {J} 1 − {J} 2 model studied with neural network quantum states},
  volume   = {100},
  issn     = {2469-9950, 2469-9969},
  url      = {https://link.aps.org/doi/10.1103/PhysRevB.100.125124},
  doi      = {10.1103/PhysRevB.100.125124},
  language = {en},
  number   = {12},
  urldate  = {2025-04-20},
  journal  = {Physical Review B},
  author   = {Choo, Kenny and Neupert, Titus and Carleo, Giuseppe},
  month    = sep,
  year     = {2019},
  pages    = {125124}
}

@article{gao_efficient_2017,
  title     = {Efficient representation of quantum many-body states with deep neural networks},
  volume    = {8},
  copyright = {2017 The Author(s)},
  issn      = {2041-1723},
  url       = {https://www.nature.com/articles/s41467-017-00705-2},
  doi       = {10.1038/s41467-017-00705-2},
  language  = {en},
  number    = {1},
  urldate   = {2025-04-20},
  journal   = {Nature Communications},
  publisher = {Nature Publishing Group},
  author    = {Gao, Xun and Duan, Lu-Ming},
  month     = sep,
  year      = {2017},
  pages     = {662}
}

@article{nomura_restricted_2017,
  title     = {Restricted {Boltzmann} machine learning for solving strongly correlated quantum systems},
  volume    = {96},
  copyright = {https://link.aps.org/licenses/aps-default-license},
  issn      = {2469-9950, 2469-9969},
  url       = {https://link.aps.org/doi/10.1103/PhysRevB.96.205152},
  doi       = {10.1103/PhysRevB.96.205152},
  language  = {en},
  number    = {20},
  urldate   = {2025-04-20},
  journal   = {Physical Review B},
  author    = {Nomura, Yusuke and Darmawan, Andrew S. and Yamaji, Youhei and Imada, Masatoshi},
  month     = nov,
  year      = {2017},
  pages     = {205152}
}

@article{nomura_helping_2021,
  title     = {Helping restricted {Boltzmann} machines with quantum-state representation by restoring symmetry},
  volume    = {33},
  issn      = {0953-8984},
  url       = {https://dx.doi.org/10.1088/1361-648X/abe268},
  doi       = {10.1088/1361-648X/abe268},
  language  = {en},
  number    = {17},
  urldate   = {2025-04-20},
  journal   = {Journal of Physics: Condensed Matter},
  publisher = {IOP Publishing},
  author    = {Nomura, Yusuke},
  month     = apr,
  year      = {2021},
  pages     = {174003}
}

@article{deng_quantum_2017,
  title     = {Quantum {Entanglement} in {Neural} {Network} {States}},
  volume    = {7},
  copyright = {https://creativecommons.org/licenses/by/4.0/},
  issn      = {2160-3308},
  url       = {http://link.aps.org/doi/10.1103/PhysRevX.7.021021},
  doi       = {10.1103/PhysRevX.7.021021},
  language  = {en},
  number    = {2},
  urldate   = {2025-04-20},
  journal   = {Physical Review X},
  author    = {Deng, Dong-Ling and Li, Xiaopeng and Das Sarma, S.},
  month     = may,
  year      = {2017},
  pages     = {021021}
}

@article{vieijra_restricted_2020,
  title    = {Restricted {Boltzmann} {Machines} for {Quantum} {States} with {Non}-{Abelian} or {Anyonic} {Symmetries}},
  volume   = {124},
  issn     = {0031-9007, 1079-7114},
  url      = {https://link.aps.org/doi/10.1103/PhysRevLett.124.097201},
  doi      = {10.1103/PhysRevLett.124.097201},
  language = {en},
  number   = {9},
  urldate  = {2025-04-20},
  journal  = {Physical Review Letters},
  author   = {Vieijra, Tom and Casert, Corneel and Nys, Jannes and De Neve, Wesley and Haegeman, Jutho and Ryckebusch, Jan and Verstraete, Frank},
  month    = mar,
  year     = {2020},
  pages    = {097201}
}

@article{machaczek_neural_2025,
  title    = {Neural quantum state study of fracton models},
  volume   = {18},
  issn     = {2542-4653},
  url      = {https://scipost.org/10.21468/SciPostPhys.18.3.112},
  doi      = {10.21468/SciPostPhys.18.3.112},
  language = {en},
  number   = {3},
  urldate  = {2025-04-20},
  journal  = {SciPost Physics},
  author   = {Machaczek, Marc and Pollet, Lode and Liu, Ke},
  month    = mar,
  year     = {2025},
  pages    = {112}
}

@article{duric_spin-_2025,
  title      = {Spin- 1 / 2 {Kagome} {Heisenberg} {Antiferromagnet}: {Machine} {Learning} {Discovery} of the {Spinon} {Pair}-{Density}-{Wave} {Ground} {State}},
  volume     = {15},
  issn       = {2160-3308},
  shorttitle = {Spin- 1 / 2 {Kagome} {Heisenberg} {Antiferromagnet}},
  url        = {https://link.aps.org/doi/10.1103/PhysRevX.15.011047},
  doi        = {10.1103/PhysRevX.15.011047},
  language   = {en},
  number     = {1},
  urldate    = {2025-04-20},
  journal    = {Physical Review X},
  author     = {Đurić, Tanja and Chung, Jia Hui and Yang, Bo and Sengupta, Pinaki},
  month      = mar,
  year       = {2025},
  pages      = {011047}
}

@misc{teng_solving_2024,
  title         = {Solving and visualizing fractional quantum {Hall} wavefunctions with neural network},
  url           = {http://arxiv.org/abs/2412.00618},
  howpublished  = {arXiv preprint arXiv:2412.00618},
  eprint        = {2412.00618},
  archiveprefix = {arXiv},
  urldate       = {2025-04-20},
  author        = {Teng, Yi and Dai, David D. and Fu, Liang},
  month         = nov,
  year          = {2024}
}

@misc{luo_solving_2025,
      title={Solving fractional electron states in twisted MoTe$_2$ with deep neural network}, 
      author={Di Luo and Timothy Zaklama and Liang Fu},
      year={2025},
      eprint={2503.13585},
      archivePrefix={arXiv},
      primaryClass={cond-mat.str-el},
      url={https://arxiv.org/abs/2503.13585}, 
}

@misc{li_deep_2025,
  title         = {Deep {Learning} {Sheds} {Light} on {Integer} and {Fractional} {Topological} {Insulators}},
  url           = {http://arxiv.org/abs/2503.11756},
  howpublished  = {arXiv preprint arXiv:2503.11756},
  eprint        = {2503.11756},
  archiveprefix = {arXiv},
  urldate       = {2025-04-20},
  author        = {Li, Xiang and Chen, Yixiao and Li, Bohao and Chen, Haoxiang and Wu, Fengcheng and Chen, Ji and Ren, Weiluo},
  month         = mar,
  year          = {2025}
}

@article{loh_sign_1990,
  title     = {Sign problem in the numerical simulation of many-electron systems},
  volume    = {41},
  url       = {https://link.aps.org/doi/10.1103/PhysRevB.41.9301},
  doi       = {10.1103/PhysRevB.41.9301},
  number    = {13},
  urldate   = {2025-04-20},
  journal   = {Physical Review B},
  publisher = {American Physical Society},
  author    = {Loh, E. Y. and Gubernatis, J. E. and Scalettar, R. T. and White, S. R. and Scalapino, D. J. and Sugar, R. L.},
  month     = may,
  year      = {1990},
  pages     = {9301--9307}
}

@article{henelius_sign_2000,
  title     = {Sign problem in {Monte} {Carlo} simulations of frustrated quantum spin systems},
  volume    = {62},
  url       = {https://link.aps.org/doi/10.1103/PhysRevB.62.1102},
  doi       = {10.1103/PhysRevB.62.1102},
  number    = {2},
  urldate   = {2025-04-20},
  journal   = {Physical Review B},
  publisher = {American Physical Society},
  author    = {Henelius, Patrik and Sandvik, Anders W.},
  month     = jul,
  year      = {2000},
  pages     = {1102--1113}
}

@article{schuch_computational_2007,
  title     = {Computational {Complexity} of {Projected} {Entangled} {Pair} {States}},
  volume    = {98},
  url       = {https://link.aps.org/doi/10.1103/PhysRevLett.98.140506},
  doi       = {10.1103/PhysRevLett.98.140506},
  number    = {14},
  urldate   = {2025-04-20},
  journal   = {Physical Review Letters},
  publisher = {American Physical Society},
  author    = {Schuch, Norbert and Wolf, Michael M. and Verstraete, Frank and Cirac, J. Ignacio},
  month     = apr,
  year      = {2007},
  pages     = {140506}
}

@article{tagliacozzo_simulation_2009,
  title     = {Simulation of two-dimensional quantum systems using a tree tensor network that exploits the entropic area law},
  volume    = {80},
  copyright = {http://link.aps.org/licenses/aps-default-license},
  issn      = {1098-0121, 1550-235X},
  url       = {https://link.aps.org/doi/10.1103/PhysRevB.80.235127},
  doi       = {10.1103/PhysRevB.80.235127},
  language  = {en},
  number    = {23},
  urldate   = {2025-04-20},
  journal   = {Physical Review B},
  author    = {Tagliacozzo, L. and Evenbly, G. and Vidal, G.},
  month     = dec,
  year      = {2009},
  pages     = {235127}
}

@article{sandvik_computational_2010,
  title     = {Computational {Studies} of {Quantum} {Spin} {Systems}},
  volume    = {1297},
  issn      = {0094-243X},
  url       = {https://pubs.aip.org/aip/acp/article/1297/1/135/854814/Computational-Studies-of-Quantum-Spin-Systems},
  doi       = {10.1063/1.3518900},
  language  = {en},
  number    = {1},
  urldate   = {2025-05-25},
  journal   = {AIP Conference Proceedings},
  publisher = {AIP Publishing},
  author    = {Sandvik, Anders W.},
  month     = nov,
  year      = {2010},
  pages     = {135--338}
}

@article{aaronson_shadow_2020,
  title     = {Shadow {Tomography} of {Quantum} {States}},
  volume    = {49},
  issn      = {0097-5397},
  url       = {https://epubs.siam.org/doi/10.1137/18M120275X},
  doi       = {10.1137/18M120275X},
  number    = {5},
  urldate   = {2025-05-25},
  journal   = {SIAM Journal on Computing},
  publisher = {Society for Industrial and Applied Mathematics},
  author    = {Aaronson, Scott},
  month     = jan,
  year      = {2020},
  pages     = {STOC18--368}
}

@article{white_density_1992,
  title     = {Density matrix formulation for quantum renormalization groups},
  volume    = {69},
  copyright = {http://link.aps.org/licenses/aps-default-license},
  issn      = {0031-9007},
  url       = {https://link.aps.org/doi/10.1103/PhysRevLett.69.2863},
  doi       = {10.1103/PhysRevLett.69.2863},
  language  = {en},
  number    = {19},
  urldate   = {2025-05-25},
  journal   = {Physical Review Letters},
  author    = {White, Steven R.},
  month     = nov,
  year      = {1992},
  pages     = {2863--2866}
}

@article{hornik_multilayer_1989,
  title    = {Multilayer feedforward networks are universal approximators},
  volume   = {2},
  issn     = {0893-6080},
  url      = {https://www.sciencedirect.com/science/article/pii/0893608089900208},
  doi      = {10.1016/0893-6080(89)90020-8},
  number   = {5},
  urldate  = {2025-05-25},
  journal  = {Neural Networks},
  author   = {Hornik, Kurt and Stinchcombe, Maxwell and White, Halbert},
  month    = jan,
  year     = {1989},
  pages    = {359--366}
}

@article{sharir_neural_2022,
  title    = {Neural tensor contractions and the expressive power of deep neural quantum states},
  volume   = {106},
  issn     = {2469-9950, 2469-9969},
  url      = {https://link.aps.org/doi/10.1103/PhysRevB.106.205136},
  doi      = {10.1103/PhysRevB.106.205136},
  language = {en},
  number   = {20},
  urldate  = {2025-05-25},
  journal  = {Physical Review B},
  author   = {Sharir, Or and Shashua, Amnon and Carleo, Giuseppe},
  month    = nov,
  year     = {2022},
  pages    = {205136}
}

@article{huang_neural_2021,
  title    = {Neural {Network} {Representation} of {Tensor} {Network} and {Chiral} {States}},
  volume   = {127},
  issn     = {0031-9007, 1079-7114},
  url      = {https://link.aps.org/doi/10.1103/PhysRevLett.127.170601},
  doi      = {10.1103/PhysRevLett.127.170601},
  language = {en},
  number   = {17},
  urldate  = {2025-05-25},
  journal  = {Physical Review Letters},
  author   = {Huang, Yichen and Moore, Joel E.},
  month    = oct,
  year     = {2021},
  pages    = {170601}
}

@article{wu_tensor-network_2023,
  title    = {From tensor-network quantum states to tensorial recurrent neural networks},
  volume   = {5},
  issn     = {2643-1564},
  url      = {https://link.aps.org/doi/10.1103/PhysRevResearch.5.L032001},
  doi      = {10.1103/PhysRevResearch.5.L032001},
  language = {en},
  number   = {3},
  urldate  = {2025-05-25},
  journal  = {Physical Review Research},
  author   = {Wu, Dian and Rossi, Riccardo and Vicentini, Filippo and Carleo, Giuseppe},
  month    = jul,
  year     = {2023},
  pages    = {L032001}
}

@article{roth_high-accuracy_2023,
  title    = {High-accuracy variational {Monte} {Carlo} for frustrated magnets with deep neural networks},
  volume   = {108},
  issn     = {2469-9950, 2469-9969},
  url      = {https://link.aps.org/doi/10.1103/PhysRevB.108.054410},
  doi      = {10.1103/PhysRevB.108.054410},
  language = {en},
  number   = {5},
  urldate  = {2025-05-25},
  journal  = {Physical Review B},
  author   = {Roth, Christopher and Szabó, Attila and MacDonald, Allan H.},
  month    = aug,
  year     = {2023},
  pages    = {054410}
}

@article{zheng_restricted_2019,
  title    = {Restricted {Boltzmann} machines and matrix product states of one-dimensional translationally invariant stabilizer codes},
  volume   = {99},
  issn     = {2469-9950, 2469-9969},
  url      = {https://link.aps.org/doi/10.1103/PhysRevB.99.155129},
  doi      = {10.1103/PhysRevB.99.155129},
  language = {en},
  number   = {15},
  urldate  = {2025-05-25},
  journal  = {Physical Review B},
  author   = {Zheng, Yunqin and He, Huan and Regnault, Nicolas and Bernevig, B. Andrei},
  month    = apr,
  year     = {2019},
  pages    = {155129}
}

@article{liang_solving_2018,
  title    = {Solving frustrated quantum many-particle models with convolutional neural networks},
  volume   = {98},
  issn     = {2469-9950, 2469-9969},
  url      = {https://link.aps.org/doi/10.1103/PhysRevB.98.104426},
  doi      = {10.1103/PhysRevB.98.104426},
  language = {en},
  number   = {10},
  urldate  = {2025-05-25},
  journal  = {Physical Review B},
  author   = {Liang, Xiao and Liu, Wen-Yuan and Lin, Pei-Ze and Guo, Guang-Can and Zhang, Yong-Sheng and He, Lixin},
  month    = sep,
  year     = {2018},
  pages    = {104426}
}

@article{viteritti_transformer_2023,
  title    = {Transformer {Variational} {Wave} {Functions} for {Frustrated} {Quantum} {Spin} {Systems}},
  volume   = {130},
  issn     = {0031-9007, 1079-7114},
  url      = {https://link.aps.org/doi/10.1103/PhysRevLett.130.236401},
  doi      = {10.1103/PhysRevLett.130.236401},
  language = {en},
  number   = {23},
  urldate  = {2025-05-25},
  journal  = {Physical Review Letters},
  author   = {Viteritti, Luciano Loris and Rende, Riccardo and Becca, Federico},
  month    = jun,
  year     = {2023},
  pages    = {236401}
}

@article{qian_describing_2025,
  title    = {Describing {Landau} {Level} {Mixing} in {Fractional} {Quantum} {Hall} {States} with {Deep} {Learning}},
  volume   = {134},
  issn     = {0031-9007, 1079-7114},
  url      = {https://link.aps.org/doi/10.1103/PhysRevLett.134.176503},
  doi      = {10.1103/PhysRevLett.134.176503},
  language = {en},
  number   = {17},
  urldate  = {2025-05-25},
  journal  = {Physical Review Letters},
  author   = {Qian, Yubing and Zhao, Tongzhou and Zhang, Jianxiao and Xiang, Tao and Li, Xiang and Chen, Ji},
  month    = apr,
  year     = {2025},
  pages    = {176503}
}

@inproceedings{klambauer_self-normalizing_2017,
  title     = {Self-{Normalizing} {Neural} {Networks}},
  volume    = {30},
  url       = {https://proceedings.neurips.cc/paper_files/paper/2017/hash/5d44ee6f2c3f71b73125876103c8f6c4-Abstract.html},
  urldate   = {2025-05-26},
  booktitle = {Advances in {Neural} {Information} {Processing} {Systems}},
  publisher = {Curran Associates, Inc.},
  author    = {Klambauer, Günter and Unterthiner, Thomas and Mayr, Andreas and Hochreiter, Sepp},
  year      = {2017}
}

@misc{kenyon2009lecturesdimers,
      title={Lectures on Dimers}, 
      author={Richard Kenyon},
      year={2009},
      eprint={0910.3129},
      archivePrefix={arXiv},
      primaryClass={math.PR},
      url={https://arxiv.org/abs/0910.3129}, 
}

@article{allegra_exact_2015,
  title      = {Exact solution of the 2d dimer model: {Corner} free energy, correlation functions and combinatorics},
  volume     = {894},
  issn       = {05503213},
  shorttitle = {Exact solution of the 2d dimer model},
  url        = {https://linkinghub.elsevier.com/retrieve/pii/S0550321315001042},
  doi        = {10.1016/j.nuclphysb.2015.03.022},
  language   = {en},
  urldate    = {2025-05-27},
  journal    = {Nuclear Physics B},
  author     = {Allegra, Nicolas},
  month      = may,
  year       = {2015},
  pages      = {685--732}
}

@article{kenyon_double-dimers_2016,
  title    = {Double-dimers, the {Ising} model and the hexahedron recurrence},
  volume   = {137},
  issn     = {00973165},
  url      = {https://linkinghub.elsevier.com/retrieve/pii/S0097316515000928},
  doi      = {10.1016/j.jcta.2015.07.005},
  language = {en},
  urldate  = {2025-05-27},
  journal  = {Journal of Combinatorial Theory, Series A},
  author   = {Kenyon, Richard and Pemantle, Robin},
  month    = jan,
  year     = {2016},
  pages    = {27--63}
}

@article{jenne_combinatorics_2021,
  title    = {Combinatorics of the double-dimer model},
  volume   = {392},
  issn     = {00018708},
  url      = {https://linkinghub.elsevier.com/retrieve/pii/S0001870821003911},
  doi      = {10.1016/j.aim.2021.107952},
  language = {en},
  urldate  = {2025-05-27},
  journal  = {Advances in Mathematics},
  author   = {Jenne, Helen},
  month    = dec,
  year     = {2021},
  pages    = {107952}
}

@article{vicentini_netket_2022,
  title      = {{NetKet} 3: {Machine} {Learning} {Toolbox} for {Many}-{Body} {Quantum} {Systems}},
  shorttitle = {{NetKet} 3},
  url        = {https://scipost.org/10.21468/SciPostPhysCodeb.7},
  doi        = {10.21468/SciPostPhysCodeb.7},
  language   = {en},
  urldate    = {2025-05-27},
  journal    = {SciPost Physics Codebases},
  author     = {Vicentini, Filippo and Hofmann, Damian and Szabó, Attila and Wu, Dian and Roth, Christopher and Giuliani, Clemens and Pescia, Gabriel and Nys, Jannes and Vargas-Calderón, Vladimir and Astrakhantsev, Nikita and Carleo, Giuseppe},
  month      = aug,
  year       = {2022},
  pages      = {7}
}

@article{carleo_netket_2019,
  title      = {{NetKet}: {A} machine learning toolkit for many-body quantum systems},
  volume     = {10},
  issn       = {23527110},
  shorttitle = {{NetKet}},
  url        = {https://linkinghub.elsevier.com/retrieve/pii/S2352711019300974},
  doi        = {10.1016/j.softx.2019.100311},
  language   = {en},
  urldate    = {2025-05-29},
  journal    = {SoftwareX},
  author     = {Carleo, Giuseppe and Choo, Kenny and Hofmann, Damian and Smith, James E.T. and Westerhout, Tom and Alet, Fabien and Davis, Emily J. and Efthymiou, Stavros and Glasser, Ivan and Lin, Sheng-Hsuan and Mauri, Marta and Mazzola, Guglielmo and Mendl, Christian B. and Van Nieuwenburg, Evert and O’Reilly, Ossian and Théveniaut, Hugo and Torlai, Giacomo and Vicentini, Filippo and Wietek, Alexander},
  month      = jul,
  year       = {2019},
  pages      = {100311}
}

@misc{mehta2014exactmappingvariationalrenormalization,
      title={An exact mapping between the Variational Renormalization Group and Deep Learning}, 
      author={Pankaj Mehta and David J. Schwab},
      year={2014},
      eprint={1410.3831},
      archivePrefix={arXiv},
      primaryClass={stat.ML},
      url={https://arxiv.org/abs/1410.3831}, 
}

@article{viteritti_accuracy_2022,
  title    = {Accuracy of restricted {Boltzmann} machines for the one-dimensional ${J}_1-{J}_2$ {Heisenberg} model},
  volume   = {12},
  issn     = {2542-4653},
  url      = {https://scipost.org/10.21468/SciPostPhys.12.5.166},
  doi      = {10.21468/SciPostPhys.12.5.166},
  language = {en},
  number   = {5},
  urldate  = {2025-05-29},
  journal  = {SciPost Physics},
  author   = {Viteritti, Luciano Loris and Ferrari, Francesco and Becca, Federico},
  month    = may,
  year     = {2022},
  pages    = {166}
}

@article{yan_sweeping_2019,
  title    = {Sweeping cluster algorithm for quantum spin systems with strong geometric restrictions},
  volume   = {99},
  issn     = {2469-9950, 2469-9969},
  url      = {https://link.aps.org/doi/10.1103/PhysRevB.99.165135},
  doi      = {10.1103/PhysRevB.99.165135},
  language = {en},
  number   = {16},
  urldate  = {2025-05-29},
  journal  = {Physical Review B},
  author   = {Yan, Zheng and Wu, Yongzheng and Liu, Chenrong and Syljuåsen, Olav F. and Lou, Jie and Chen, Yan},
  month    = apr,
  year     = {2019},
  pages    = {165135}
}

@article{moessner_resonating_2001,
  title     = {Resonating {Valence} {Bond} {Phase} in the {Triangular} {Lattice} {Quantum} {Dimer} {Model}},
  volume    = {86},
  copyright = {http://link.aps.org/licenses/aps-default-license},
  issn      = {0031-9007, 1079-7114},
  url       = {https://link.aps.org/doi/10.1103/PhysRevLett.86.1881},
  doi       = {10.1103/PhysRevLett.86.1881},
  language  = {en},
  number    = {9},
  urldate   = {2025-09-12},
  journal   = {Physical Review Letters},
  author    = {Moessner, R. and Sondhi, S. L.},
  month     = feb,
  year      = {2001},
  pages     = {1881--1884}
}

@article{sutherland_systems_1988,
  title      = {Systems with resonating-valence-bond ground states: {Correlations} and excitations},
  volume     = {37},
  copyright  = {http://link.aps.org/licenses/aps-default-license},
  issn       = {0163-1829},
  shorttitle = {Systems with resonating-valence-bond ground states},
  url        = {https://link.aps.org/doi/10.1103/PhysRevB.37.3786},
  doi        = {10.1103/PhysRevB.37.3786},
  language   = {en},
  number     = {7},
  urldate    = {2025-09-12},
  journal    = {Physical Review B},
  author     = {Sutherland, Bill},
  month      = mar,
  year       = {1988},
  pages      = {3786--3789}
}

@article{moessner_short-ranged_2001,
  title     = {Short-ranged resonating valence bond physics, quantum dimer models, and {Ising} gauge theories},
  volume    = {65},
  copyright = {http://link.aps.org/licenses/aps-default-license},
  issn      = {0163-1829, 1095-3795},
  url       = {https://link.aps.org/doi/10.1103/PhysRevB.65.024504},
  doi       = {10.1103/PhysRevB.65.024504},
  language  = {en},
  number    = {2},
  urldate   = {2025-09-12},
  journal   = {Physical Review B},
  author    = {Moessner, R. and Sondhi, S. L. and Fradkin, Eduardo},
  month     = dec,
  year      = {2001},
  pages     = {024504}
}

@article{kasteleyn_statistics_1961,
  title    = {The statistics of dimers on a lattice: {I}. {The} number of dimer arrangements on a quadratic lattice},
  volume   = {27},
  issn     = {0031-8914},
  url      = {https://www.sciencedirect.com/science/article/pii/0031891461900635},
  doi      = {https://doi.org/10.1016/0031-8914(61)90063-5},
  number   = {12},
  journal  = {Physica},
  author   = {Kasteleyn, P. W.},
  year     = {1961},
  pages    = {1209--1225}
}

@article{nienhuis_triangular_1984,
  title    = {Triangular {SOS} models and cubic-crystal shapes},
  volume   = {17},
  issn     = {0305-4470},
  url      = {https://dx.doi.org/10.1088/0305-4470/17/18/025},
  doi      = {10.1088/0305-4470/17/18/025},
  language = {en},
  number   = {18},
  urldate  = {2025-09-12},
  journal  = {Journal of Physics A: Mathematical and General},
  author   = {Nienhuis, B. and Hilhorst, H. J. and Blote, H. W. J.},
  month    = dec,
  year     = {1984},
  pages    = {3559}
}

@misc{fan2026equivariantneuralnetworksforcefield,
      title={Equivariant Neural Networks for Force-Field Models of Lattice Systems}, 
      author={Yunhao Fan and Gia-Wei Chern},
      year={2026},
      eprint={2601.04104},
      archivePrefix={arXiv},
      primaryClass={cond-mat.str-el},
      url={https://arxiv.org/abs/2601.04104}, 
}

@misc{raikos2026variationalstudymagnetizationplateaus,
      title={Variational study of the magnetization plateaus in the spin-1/2 kagome Heisenberg antiferromagnet: an approach from vision transformer neural quantum states}, 
      author={Andreas Raikos and Sylvain Capponi and Fabien Alet},
      year={2026},
      eprint={2602.12998},
      archivePrefix={arXiv},
      primaryClass={cond-mat.str-el},
      url={https://arxiv.org/abs/2602.12998}, 
}

@misc{chen_neural_2025,
      title={Neural Network-Augmented Pfaffian Wave-functions for Scalable Simulations of Interacting Fermions}, 
      author={Ao Chen and Zhou-Quan Wan and Anirvan Sengupta and Antoine Georges and Christopher Roth},
      year={2025},
      eprint={2507.10705},
      archivePrefix={arXiv},
      primaryClass={cond-mat.str-el},
      url={https://arxiv.org/abs/2507.10705}, 
}

@article{romero_spectroscopy_2025,
  title     = {Spectroscopy of two-dimensional interacting lattice electrons using symmetry-aware neural backflow transformations},
  volume    = {8},
  copyright = {2025 The Author(s)},
  issn      = {2399-3650},
  url       = {https://www.nature.com/articles/s42005-025-01955-z},
  doi       = {10.1038/s42005-025-01955-z},
  language  = {en},
  number    = {1},
  urldate   = {2026-04-04},
  journal   = {Communications Physics},
  publisher = {Nature Publishing Group},
  author    = {Romero, Imelda and Nys, Jannes and Carleo, Giuseppe},
  month     = jan,
  year      = {2025},
  pages     = {46}
}

@misc{rende_transformer_2026,
       title={Transformer Neural-Network Quantum States for lattice models of spins and fermions: Application to the Ancilla Layer Model}, 
      author={Riccardo Rende and Alexander Nikolaenko and Luciano Loris Viteritti and Subir Sachdev and Ya-Hui Zhang},
      year={2026},
      eprint={2603.02316},
      archivePrefix={arXiv},
      primaryClass={cond-mat.str-el},
      url={https://arxiv.org/abs/2603.02316}, 
}

@misc{nazaryan_artificial_2026,
      title={Artificial Intelligence for Quantum Matter: Finding a Needle in a Haystack}, 
      author={Khachatur Nazaryan and Filippo Gaggioli and Yi Teng and Liang Fu},
      year={2026},
      eprint={2507.13322},
      archivePrefix={arXiv},
      primaryClass={cond-mat.str-el},
      url={https://arxiv.org/abs/2507.13322}, 
}

@article{teng_solving_2025,
  title    = {Solving the fractional quantum {Hall} problem with self-attention neural network},
  volume   = {111},
  issn     = {2469-9950, 2469-9969},
  url      = {https://link.aps.org/doi/10.1103/PhysRevB.111.205117},
  doi      = {10.1103/PhysRevB.111.205117},
  language = {en},
  number   = {20},
  urldate  = {2026-04-05},
  journal  = {Physical Review B},
  author   = {Teng, Yi and Dai, David D. and Fu, Liang},
  month    = may,
  year     = {2025},
  pages    = {205117}
}

@article{PhysRevLett.121.167204,
  title = {Symmetries and Many-Body Excitations with Neural-Network Quantum States},
  author = {Choo, Kenny and Carleo, Giuseppe and Regnault, Nicolas and Neupert, Titus},
  journal = {Phys. Rev. Lett.},
  volume = {121},
  issue = {16},
  pages = {167204},
  numpages = {6},
  year = {2018},
  month = {Oct},
  publisher = {American Physical Society},
  doi = {10.1103/PhysRevLett.121.167204},
  url = {https://link.aps.org/doi/10.1103/PhysRevLett.121.167204}
}

\clearpage

\onecolumngrid
\appendix
\setcounter{page}{1}
\renewcommand\thefigure{A\arabic{figure}}
\setcounter{figure}{0}

\section{p4m irreps and characters via induction from the little groups}
\label{app:p4m_irreps}
In this section, we will consider the action of the ${\rm{p4m}}{=}\mathbb{T}{\rtimes}D_4$ group on a square lattice $\mathbb{L}$ (isomorphic to $\mathbb{T}$ itself) of size $ L{\times}L$ with the goal of identifying its irreducible representations and corresponding characters.
Each element of the group can be decomposed uniquely as $(a,\alpha)$ which represents a translation by $a{\in}\mathbb{T}$ following a $D_4$ transformation $\alpha$. The action of the group on the lattice can be used to associate a group action on the set of all dimer configurations on the $\mathbb{L}$. The group acts on the set $\Sigma$ of dimer configurations $\sigma{:} \mathbb{L}{\to}\{1,2,3,4\}$ to produce a new configuration $(a,\alpha)\sigma{\in}\Sigma$ defined by 
\begin{equation}
    ((a,\alpha)\sigma)(x) = P_{\alpha}\sigma(\alpha^{-1} (x - a))\nonumber
\end{equation}
where $P$ is a permutation representation of $D_4$ on the set $\{1,2,3,4\}$. 
The resulting composition rule is $(a,\alpha )(b,\beta)=(a + \alpha b, \alpha \beta)$. 
Translation orbits split $\Sigma$ into equivalence classes $\Sigma/\mathbb{T}$. The irreducible invariant subspaces of $\mathbb{T}$ action on $\rm{Span}_{\mathbb{C}}[\Sigma]$ are labeled by $[\sigma]\in \Sigma/\mathbb{T}$ and Brillouin zone points $k{=}(k_x,k_y)$ where $k_x,k_y \in \{0, \frac{2\pi}{L}, \ldots, \frac{2\pi (L-1)}{L}\}$.\\ 

These 1D irreps are spanned by translation eigenstates $\Phi^\sigma_k\in {\rm Span}_{\mathbb{C}}[\Sigma]$:
\begin{equation}
\Phi^\sigma_k = \sum_{a\in\mathbb{T}} e^{\dot{\imath} k \cdot a} (a,e)\sigma,\nonumber
\end{equation}
where $e $ denotes the identity element in $D_4$.
The action of ${\rm{p4m}}$ on these translation eigenstates is given by
\begin{equation}
(b,\beta)\Phi_{k}^{\sigma}\left(x\right)=e^{- \dot{\imath} \beta\left(k\right)\cdot b}\Phi_{\beta\left(k\right)}^{\beta\sigma}\nonumber
\end{equation}
where $\beta(k)$ is the action of $D_4$ on the momentum $k$ and $\beta\sigma$ is short for $(0,\beta)\sigma$. \\

In what follows, we will represent $90$ degree counter clockwise rotation by $r$ and mirror reflection about the $x$-axis, $y$-axis, diagonal ($x{=}y$) and anti-diagonal($x{=}-y$) with $m_x$, $m_y$, $m_{\rm{diag}}$ and $m_{\rm{anti\mbox{-}diag}}$, respectively.

The irreps of the $\rm{p4m}$ group can be labeled by the $k$ values modulo $D_4$ and a little group label. For a given of $k$ mod($D_4$), the different irreps of $\rm{p4m}$ are labeled by the little group (stabilizer in $D_4$ of $k$) irreps. We have the following cases:

\begin{enumerate}
\item Generic momenta $k=(k_x,k_y)$ with $k_x,k_y\not \in\{0,\pi\}$. There are $(L-2)(L-4)/8$ such momenta modulo $D4$.
$D_4$ acts freely on such $k$ values, generating an orbit of size 8. The little group is trivial and the ${\rm p4m}$ irrep is 8-dimensional, spanned by $\{\Phi_{\beta\left(k\right)},\;\beta{\in}D_4\}$. The characters are given by 
\begin{equation}
\chi_{(a,\alpha)}=\begin{cases}
0 & \alpha\neq e\\
\sum_{\beta\in D_{4}}e^{-i\beta(k)\cdot a} & \alpha=e
\end{cases}  
\end{equation}
\item Points on the high-symmetry coordinate lines $\{(\pm k,0),(0,\pm k)\}$ where $k\neq \{0, \pi\}$. There are $(L-2)/2$ such momenta mod $D_4$. The little group $G_L{\subset}D_4$ of $(k,0)$ and $(0,k)$ are $\langle m_x \rangle$ and $\langle m_y \rangle$ respectively.
There are two four-dimensional irreps $A_\pm$ of $\rm p4m$ ($\pm$ associated with the eigenvalues of the $G_L$ generators) spanned by
\begin{equation}
\{ \Psi_{g(k,0)}^{g\sigma} \pm\Psi^{gm_x {\sigma}}_{g(k,0)}\text{ where } g\in D_4/\langle { m}_x \rangle\}
\end{equation}
where action of coset element on states and momenta can be interpreted as action by any chosen element in that coset.
The characters are given by
\begin{equation}
\chi^\pm_{(a,\alpha)}=\begin{cases}
  \sum_{g\in D_4/\langle m_x \rangle} e^{-i g(k,0)\cdot a} & \alpha{=}e\\
  \pm \sum_{g\in {\rm Conj. Class}({\alpha})} e^{-i g(k,0)\cdot a} & \alpha\in\{m_x,m_y\}\\
0 & \text{otherwise}
\end{cases}
\end{equation}
For simplicity of presentation of results, it will be convenient to label $A_+$ and $A_-$ as $A_1$ and $A_2$, respectively.

\item High-symmetry diagonal lines $\{(\pm k,\pm k),(\pm k,\mp k)\}$ where $k\neq \{0, \pi \}$. There are $(L-2)/2$ such momenta modulo $D_4$. The little groups $G_L\subset D_4$ of $(k,k)$ and $(k,-k)$ are $\langle m_{\rm diag} \rangle$ and $\langle m_{\rm antidiag} \rangle$ respectively.
There are two four-dimensional irreps $A_\pm$ (determined by the eigenvalue of the little group generator) spanned by
\begin{equation}
\{ \Psi_{g(k,k)}^{g\sigma} \pm\Psi^{gm_{\rm diag} {\sigma}}_{g(k,k)}\text{ where } g\in D_4/\langle { m}_{\rm diag} \rangle\}
\end{equation}
The characters are
\begin{equation}
\chi_{(a,\alpha)}=\begin{cases}
  \sum_{g\in D_4/\langle m_{\rm diag} \rangle} e^{-i g(k,k)\cdot a} & \alpha=e\\
  \pm \sum_{g\in {\rm Conj. Class}({\alpha})} e^{-i g(k,k)\cdot a} & \alpha\in \{m_{\rm diag},m_{\rm anti\mbox{-}diag}\}\\
  0 & \text{otherwise}
\end{cases}
\end{equation}
As in the previous case, for simplicity of presentation of results, it will be convenient to label $A_+$ and $A_-$ as $A_1$ and $A_2$, respectively.

\item High-symmmetry point $k=(0,0)$. 
The little group is now the full $D_4$ group.
Translation acts trivially on the basis states $\Phi_{(0,0)}^\sigma$. $D_4$ acts as a permutation of the eight states
\[\{\Phi_{(0,0)}^{g\sigma} \text{ where } g\in D_4\}.\]

The irreps of ${\rm p4m}$ arise directly from irreps of $D_4$ (four 1D irreps and one 2D irrep). The characters are given by
\begin{equation}
\chi_{(a,\alpha)}=\chi^{D_4}_{\alpha}
\end{equation}
where $\chi^{D_4}_{\alpha}$ is the character of any of the five possible $D_4$ irreps. These irreps by convention are labeled $A_1,A_2,B_1,B_2$ (1D irrep), and $E$ (2D irrep). 
States in $A_1$ are invariant under $D_4$, states in $A_2$ flips sign under all mirror reflections, states in $B_1$ flip sign under $r$ ($90^o$ rotation) and diagonal mirrors, and states in $B_2$ flip under $r$ and mirror reflections about $x$ and $y$ axes. $E$ is a two-dimensional representation.

\item High-symmetry point $k=(\pi,\pi)$. Translations act as sublattice parity measurements, in other words, $(a,e)\Phi_{(\pi,\pi)}^\sigma = (-1)^{a_x+a_y}\Phi_{(\pi,\pi)}^\sigma$. Note that in our convention, where the origin $(0,0)$ about which the $D_4$ elements act is a lattice point (not a plaquette center), $D_4$ elements do not change the sublattice parity. Therefore $(a,\alpha)\Phi_{(\pi,\pi)}^\sigma=\left(-1\right)^{a_{x}+a_{y}}\Phi_{(\pi,\pi)}^{\alpha\sigma}$. The irreps of ${\rm{p4m}}$ are again related to the irreps of $D_4$ with characters
\begin{equation}
\chi_{(a,\alpha)}=\left(-1\right)^{a_{x}+a_{y}}\chi^{D_4}_{\alpha}
\end{equation}
As in the previous case, the representations can be labeled by $A_1,A_2,B_1,B_2$ and $E$.

\item High-symmetry points $k=\{(\pi,0),(0,\pi)\}$. There is only one such momentum modulo  $D_4$.
The little group now is the order $4$ group $G_L {=} D_2 {=} \langle m_x,m_y \rangle=\{e,m_x,m_y,m_x m_y{=}r^2\}$.  
There are $4$ distinct irreps of ${\rm p4m}$. All are 2D irreps. 
They are labeled by the eigenvalues $s_x,s_y=\pm 1$ of the generators $m_x,m_y$. The 4-dimensional irrep spaces for given $s_x,s_y$ are spanned by

\begin{equation}
\begin{aligned}
\Bigl\{
&\Phi^{g\sigma}_{g(\pi,0)}
+ s_x \Phi^{g m_x \sigma}_{g(\pi,0)}
+ s_y \Phi^{g m_y \sigma}_{g(\pi,0)}
+ s_x s_y \Phi^{g m_x m_y \sigma}_{g(\pi,0)}
\text{with } g \in D_4 / D_2 = \{ e, r \}\Bigr\}.
\end{aligned}
\end{equation}
The characters are given by
\begin{equation}
\chi_{(a,\alpha)}=\begin{cases}(-1)^{a_x} + (-1)^{a_y} & \alpha=e\\
(-1)^{a_x}s_x + (-1)^{a_y}s_y & \alpha=m_x\\
(-1)^{a_x}s_y + (-1)^{a_y}s_x & \alpha=m_y\\
((-1)^{a_x} + (-1)^{a_y})s_x s_y & \alpha=m_x m_y\\
0 & \text{otherwise}
\end{cases}
\end{equation}

The representation with $s_x,s_y=+1$ and $s_x,s_y=-1$ will be labeled $A_1$ and $A_2$ respectively. The representations labeled $s_x=+1,s_y=-1$ and $s_x=-1,s_y=1$ are labeled $p_{\rm{long}}$ and $p_{\rm{trans}}$ respectively.

\item Coordinate axes $\{(\pi,\pm k),(\pm k,\pi)\}$. There are $(L-2)/2$ such momenta modulo $D_4$. The irreps are analogous to the coordinate axes case. $D_4$ orbit is $4$ dimensional and the little groups at $(\pi,\pm k)$ and $(\pi,\pm k)$ are  $\langle m_y\rangle$ and $\langle m_x\rangle$.

\end{enumerate} 

\clearpage
\section{Ordered states and symmetries}
\label{app:columnar_plaquette_state_illustration}
\begin{figure}[h]
    \includegraphics[width=0.8\columnwidth]{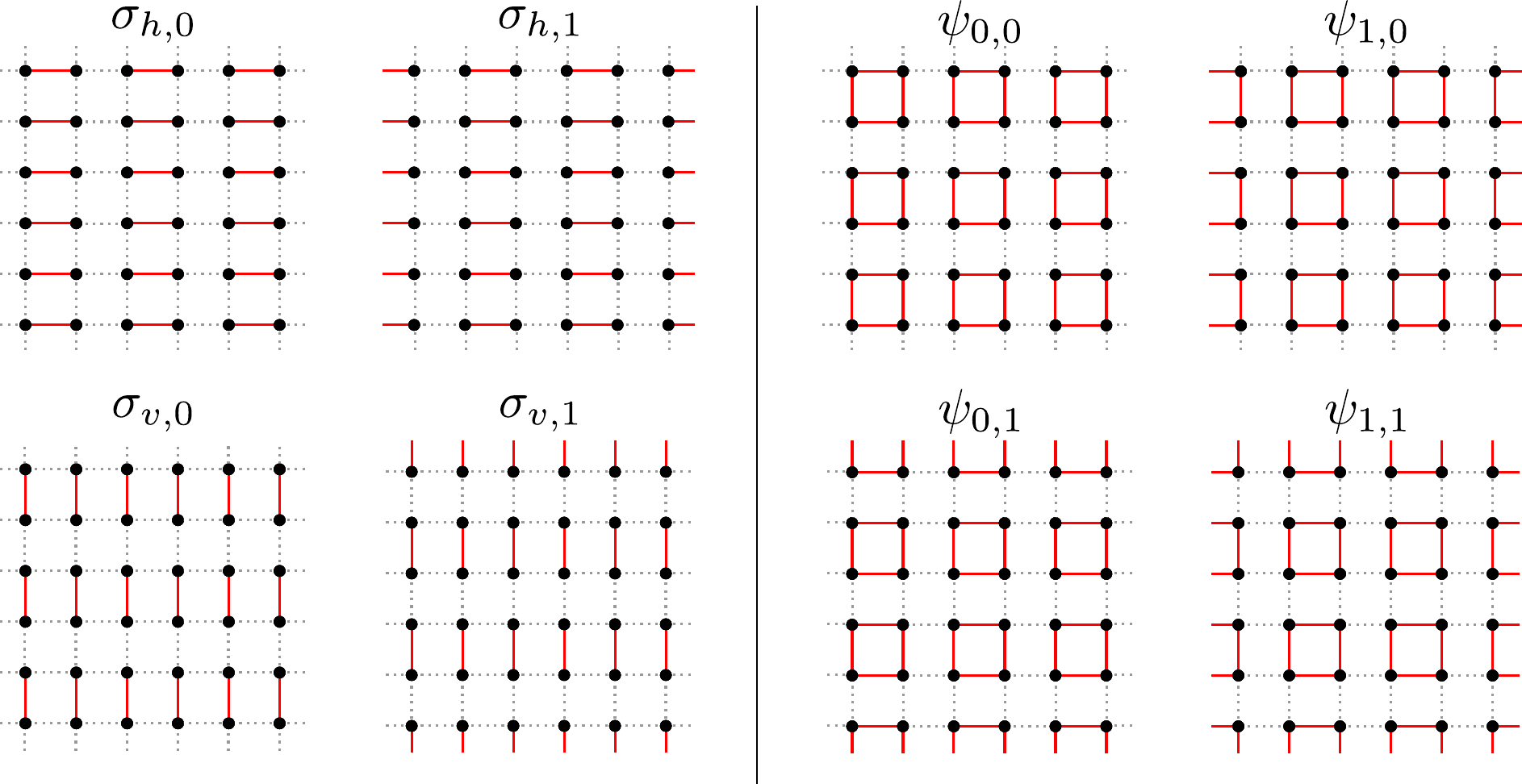}
    \caption{Schematic illustration of the four columnar states (left) and the four plaquette states (right). The red lines represent average dimer probability ($1$ for the ideal columnar state and $0.5$ for the ideal plaquette state).
    \label{fig:dimer_configs}}
\end{figure}

\paragraph*{Columnar ordered states}: The set of the four columnar-ordered states forms the following irreps 
\begin{align}
A_1, k=(0,0)&: \;\sigma_{h,0}+\sigma_{h,1}+\sigma_{v,0}+\sigma_{v,1}\nonumber\\ 
B_1, k=(0,0)&: \;\sigma_{h,0}+\sigma_{h,1}-\sigma_{v,0}-\sigma_{v,1}\nonumber\\
p_{\rm long},  k\in\{(0,\pi),(\pi,0)\}&: \; {\rm Span}_{\mathbb{C}}\left[ \sigma_{h,0}-\sigma_{h,1},\sigma_{v,0}-\sigma_{v,1} \right]
\end{align}
Here $\sigma_{h,0}$ and $\sigma_{h,1}=((1,0),e)\sigma_{h,0}$ are the two horizontally ordered columnar states and $\sigma_{v,0}=((0,0),r)\sigma_{h,0}$ and $\sigma_{v,1}=((0,1),r)\sigma_{h,0}$ are the two vertically ordered columnar states (Fig.~\ref{fig:dimer_configs}). Columnar ordered phase occurs when these states or dressed forms of these become degenerate. Thus columnar ordered phase is characterized by degenerate ground states in the $A_1, k=(0,0)$, $B_1, k=(0,0)$ and $p_{\rm long}$ sectors.\\

\paragraph*{Plaquette states}:  The space spanned by the four plaquette ordered states (Fig.~\ref{fig:dimer_configs}) $\psi_{0,0}$, $\psi_{1,0}=((1,0),e)\psi_{0,0}$, $\psi_{0,1}=((0,1),e)\psi_{0,0}$ and $\psi_{1,1}=((1,1),e)\psi_{0,0}$ decomposes into
\begin{align}
A1, k=(0,0)&: \psi_{0,0}+\psi_{1,0}+\psi_{0,1}+\psi_{1,1}\nonumber\\
B_2, k=(\pi,\pi)&: \psi_{0,0}-\psi_{1,0}-\psi_{0,1}+\psi_{1,1}\nonumber\\
p_{\rm{long}},  k\in\{(0,\pi),(\pi,0)\}&: {\rm Span}_\mathbb{C}\{ \psi_{0,0}-\psi_{1,0}+\psi_{0,1}-\psi_{1,1}, \psi_{0,0}+\psi_{1,0}-\psi_{0,1}-\psi_{1,1} \}.\nonumber
\end{align}
Plaquette ordered phase occurs when these states or dressed forms of these become degenerate. Thus plaquette ordered phase is characterized by degenerate ground states in the $A_1, k=(0,0)$, $B_2, k=(\pi,\pi)$ and $p_{\rm long}$ (2D) sectors.

Mixed state is associated with degenerate ground states in the sectors $A_1, k=(0,0)$, $B_1, k=(0,0)$, $B_2, k=(\pi,\pi)$ and $p_{\rm long}$ (2D).

\clearpage
\section{Benchmark with ED} 
\label{app:appendix_ED_L8}
\begin{figure}[h]
	\centering
    \includegraphics[width=0.5\columnwidth]{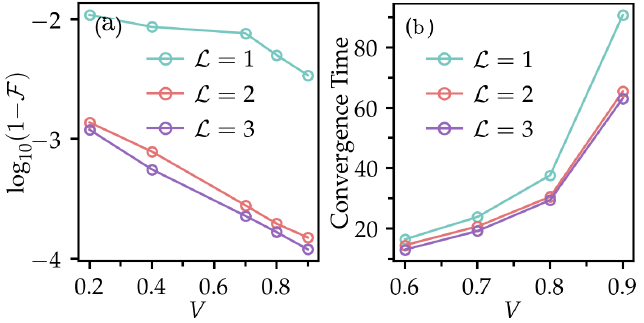}
    \caption{(a) Comparison of $\psi(\sigma)$ from GCNN and ED at $V{=}0.8$. 
    (b) Infidelity $1{-}\mathcal{F}$ vs $V$.
    (c) $\text{log}_{10}\Delta E $ vs iterations at $V{=}0.9$. 
    (d) Convergence time scale vs $V$. All calculations here are done for $L{=}8$.
    \label{fig:ed_l8}}
\end{figure}

In Fig.~\ref{fig:ed_l8}, we provide additional convergence data for the GCNN ground state of an $L{=}8$ system (Hilbert space dimension $311{,}853{,}312$) at multiple values of $V$. The GCNN performs optimization within the 628,931 dimensional ${\rm p4m}$-invariant subspace as mentioned in the main text.

\begin{figure*}
    \includegraphics[width=\textwidth]{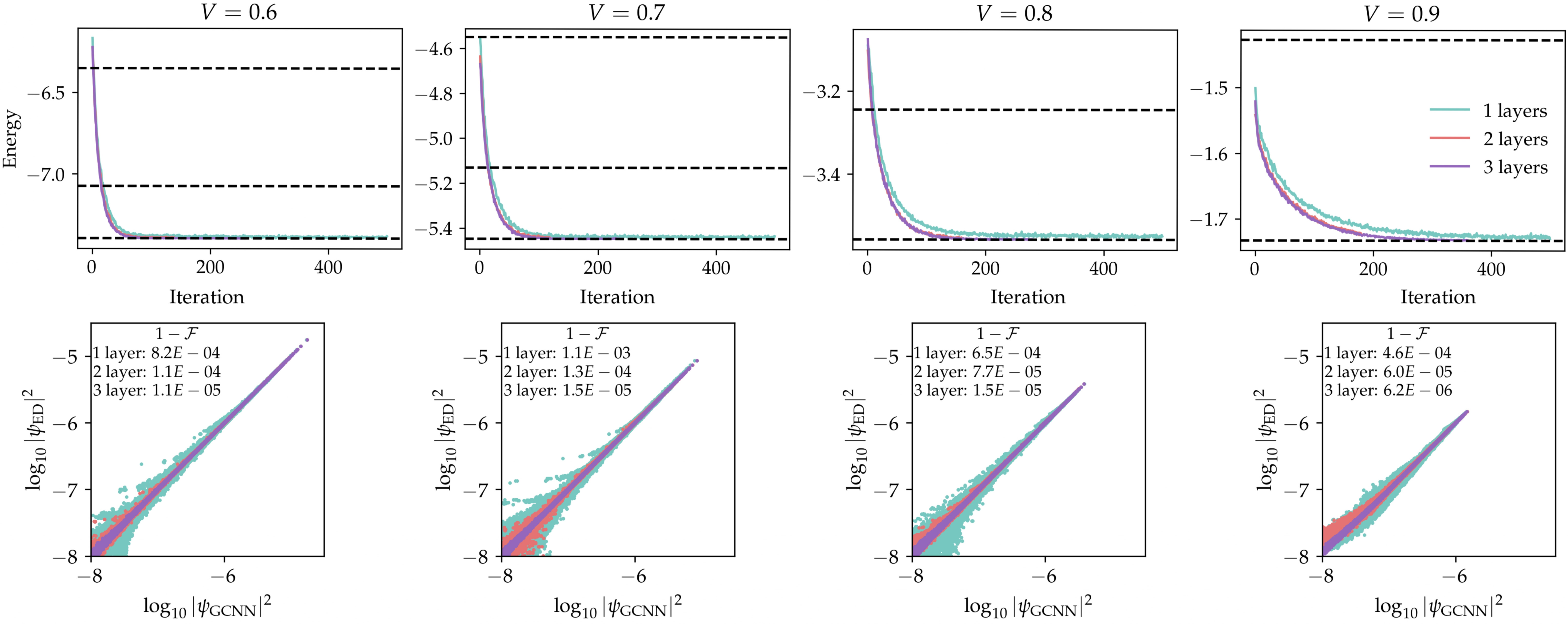}
    \caption{First row: convergence of energy with iteration for $V=0.6,0.7,0.8$ and $0.9$ on an  $8 {\times} 8$ lattice. The different black dotted lines are the energy values obtained from ED calculation for the ground state and first few excited states.\\
    Second row: Comparison between square of wavefunction amplitudes for $(0,0)$ winding sector ${\rm p4m}$-invariant states obtained from GCNN and ED calculations. 
    \label{fig:L8_ener_and_error_coeffs}} 
\end{figure*}

%
To understand the symmetry structure of this subspace, we note that the Hamiltonian commutes with the winding numbers. The $H_{\rm QDM}$ commutes with the winding number ${(W_x,W_y)}$ where $W_x{\coloneqq}\sum_{\vec{r} \in \mathcal{C}_y} (-1)^{r_y} n_x(\vec{r})$ where $r_y$ is the $y$-coordinate of the lattice site $\vec{r}$, $\mathcal{C}_y$ are the lattice sites at a fixed $r_x$, and $n_x(\vec{r}){=}\delta_{\sigma_{\vec{r},+\hat{x}}}$. Analogous definition applies to $W_y$ with $x$ and $y$ interchanged.
The ground state is in the $(W_x,W_y){=}(0,0)$ sector for $V{<}1$. The GCNN ground state's components in the $W{\neq }0$ sectors (dim $ 1.58\times 10^8$) are negligible ($\langle \psi_{\rm GCNN}|\Pi_{(W_x,W_y)\neq (0,0)}|\psi_{\rm GCNN}\rangle{\approx}10^{-5}$ at $V{=}0.8$) though the winding number constraint is not imposed on the GCNN.
As the system approaches the critical RK point, the fidelity $\mathcal{F} = \left| \langle \psi_{\text{ED}} | \psi_{\text{GCNN}} \rangle \right|^2$ between the GCNN and exact ground states improves, as shown in Fig.~\ref{fig:ed_l8}(a). 
For each $V$, $\mathcal{F}$ increases  as $\mathcal{L}$ goes from $1$ to $2$, but shows little gain from $\mathcal{L}{=}2$ to $3$. 
  Convergence time-scale in iterations (Fig.~\ref{fig:ed_l8}(b)) increases with $V$ as RK point is approached but at the RK point, where  $\psi(\sigma)$ is a constant, GCNN optimizes in $\sim 10$ iterations.
We extend our results that are shown in Fig.~\ref{fig:ed_l8} for different $V$ values ($V{=}0.6, 0.7, 0.8, 0.9$) in Fig.~\ref{fig:L8_ener_and_error_coeffs}. 
We observe a nice convergence to the exact ground state (shown as the bottom-most black dotted line) for all values of $V$. 
We also show a few excited states (obtained from ED) for $L{=}8$ in Fig.~\ref{fig:L8_ener_and_error_coeffs}. Here, it is worth noting that the fluctuations in the estimated energies are much smaller than the exact energy gap. 
We observe that the convergence time increases as we move towards the RK point, which is consistent with the results in Fig.~\ref{fig:ed_l8}(d). We also compare the coefficients ($|\psi(\sigma)|^2$ obtained from GCNN and ED for the mentioned values of $V$ in the second row of Fig.~\ref{fig:L8_ener_and_error_coeffs}.

\section{Benchmark with QMC} 

\begin{figure*}[h]
	\includegraphics[width=\textwidth]{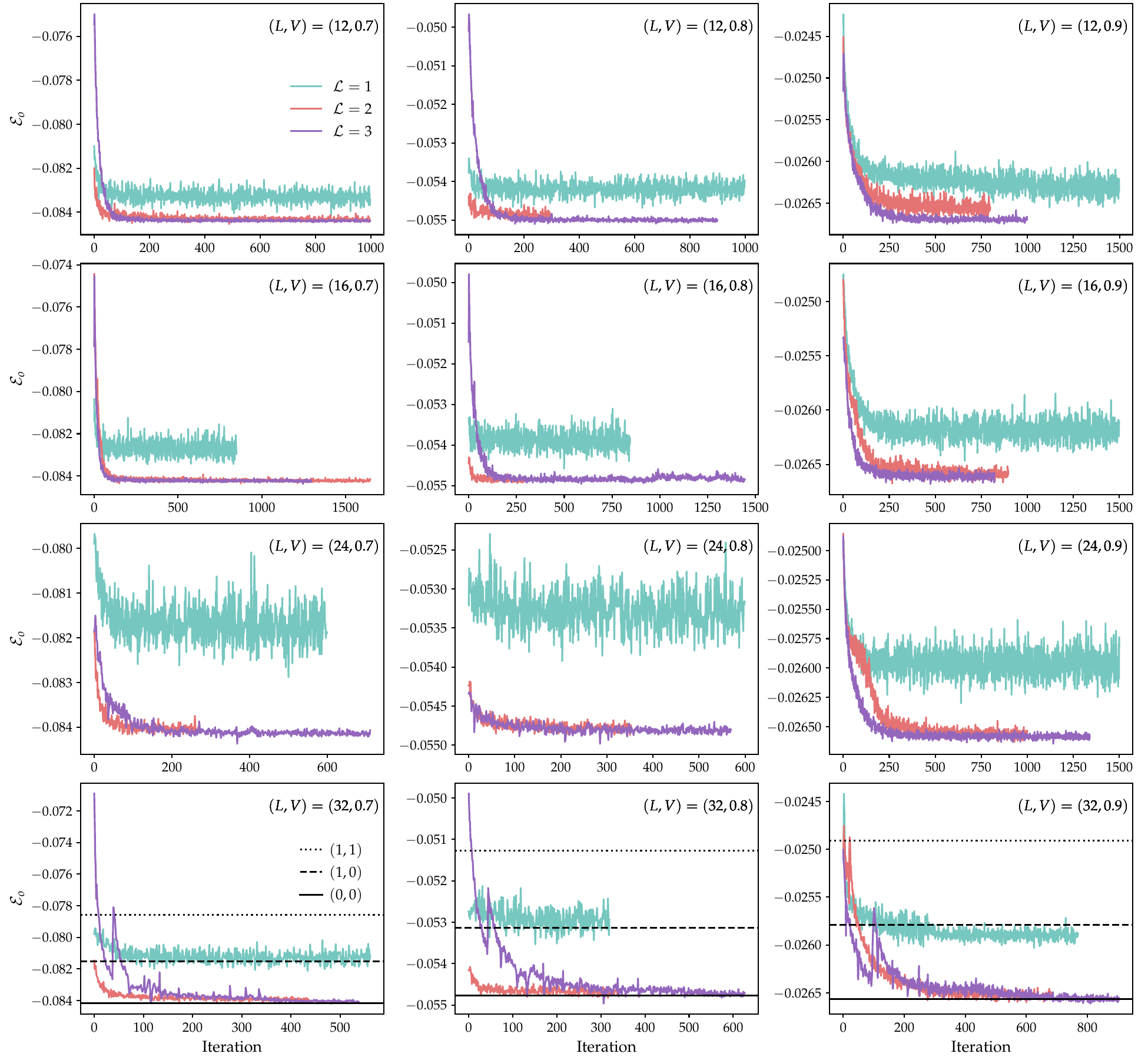}
	\caption{\label{fig:energyDensityForAppendix}Energy density as a function of iteration for $12\leq L\leq32$, $V=\{0.7,0.8,0.9\}$ and $\mathcal{L}=\{1,2,3\}$. For $L=32$, the black horizontal lines correspond to the QMC estimates~\cite{syljuasen_plaquette_2006} of energy density for different winding sectors: the solid, dashed, and dotted lines denote $(0,0)$, $(1,0)$, and $(1,1)$ winding sectors, respectively.} 
\end{figure*}
\FloatBarrier


In Fig.\ref{fig:energyDensityForAppendix}, we present energy density $\mathcal{E}_o$ (energy per plaquette) for various combinations of $(L,V,\mathcal{L})$ for $12\leq L\leq32$ and $\mathcal{L} = 1,2,3$ layers in the $\text{GCNN}$. 
We observe that $\mathcal{L}{=}2$ $\text{GCNN}$ gives results as good as $\mathcal{L}{=}3$ $\text{GCNN}$. 
To investigate the convergence behavior, we monitor the energy density as a function of iteration during the training of $\text{GCNNs}$, and find that the convergence time increases with $V$. 
Notably, once a network is trained for a given value of the interaction parameter $V$, we use its optimized parameters as the initial seed for training at a nearby $V$ value. 
We observe that this warm-start strategy leads to significantly faster convergence compared to training from a random initialization. 
We benchmark our results against QMC calculations for $L=32$~\cite{syljuasen_plaquette_2006}, and find excellent agreement between the two, confirming the reliability of our approach.

The three black horizontal lines in Fig.~\ref{fig:energyDensityForAppendix} correspond to the QMC estimates~\cite{syljuasen_plaquette_2006} of the energy density for different winding sectors: the solid, dashed, and dotted lines denote the $(0,0)$, $(1,0)$, and $(1,1)$ winding sectors, respectively.
We observe that the energy fluctuations in the $\rm GCNN$ results remain smaller than the gaps between $\rm ED$ energies of different winding sectors, ensuring that the winding sector-dependent energy differences are well resolved.

\section{Columnar order parameter distribution} 
\begin{figure}[h]
    \includegraphics[width=0.85\columnwidth]{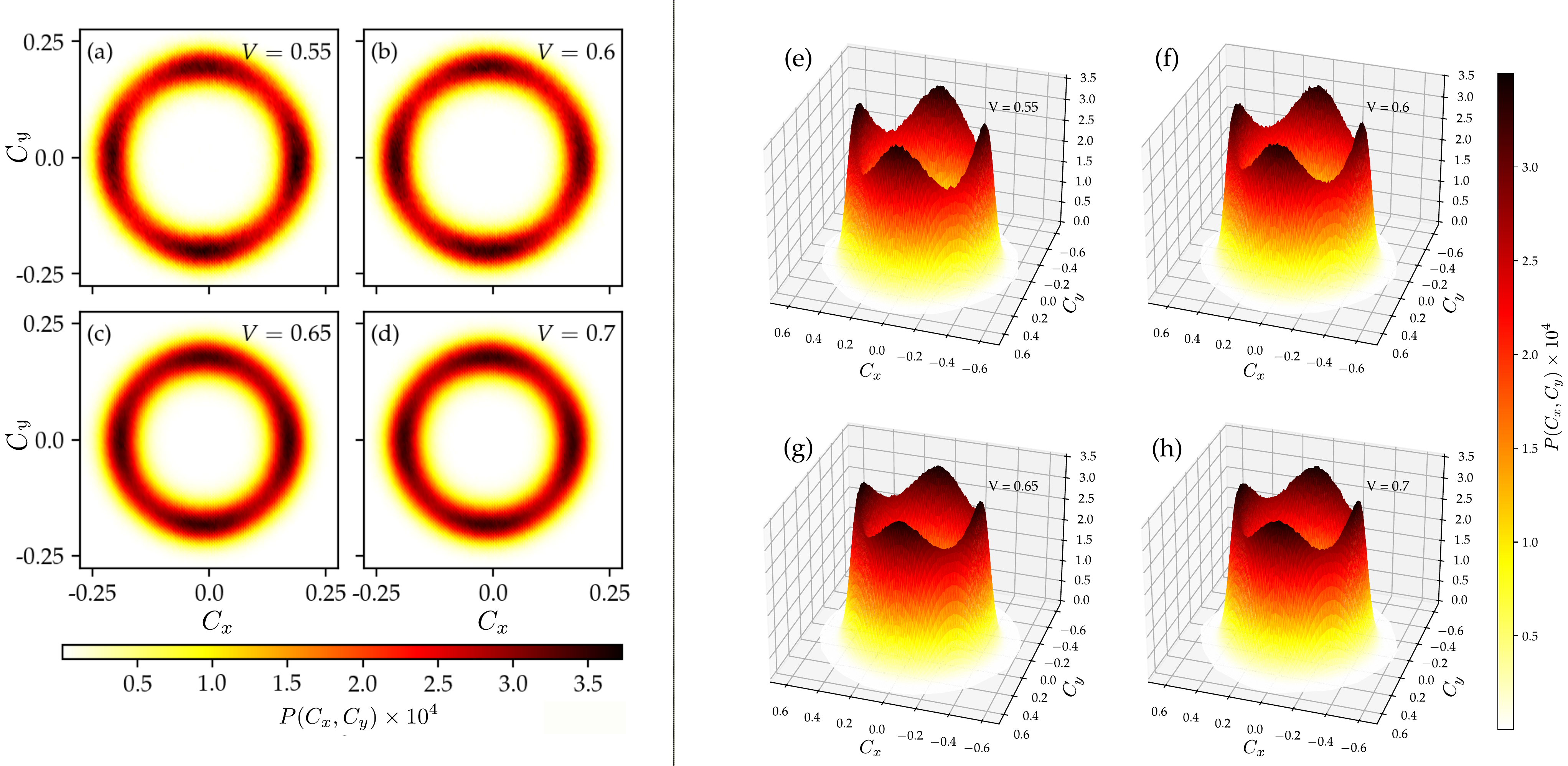}
    \caption{(a-d) $C_x,C_y$ distribution at $L{=}24$ for $V{=}0.55, 0.6, 0.65$ and $0.7$.
    (e-h) 3D surface plots for the corresponding $V$ values. The $x$- and $y$-axes denote $C_x, C_y$ respectively. The height and colour represent probability $P(C_x, C_y)$.
    \label{fig:Nx_Ny_distribution_L24}}
\end{figure}

We present the distribution of the columnar order parameter in the ground state for $L{=}24$ at different values of $V$ in Fig.~\ref{fig:Nx_Ny_distribution_L24}(a-d). 
These results demonstrate the emergence of four-fold rotational symmetry, reflected in pronounced peaks in the cardinal directions.
The angular positions of the peaks are consistent with the presence of a columnar-ordered phase, in agreement with the criterion discussed in Ref.~\cite{yan_widely_2021-1}.
The corresponding $3 \rm D$ surface plots are shown in Fig.~\ref{fig:Nx_Ny_distribution_L24}(e-f), where the peaks are easily discerned. 
The distributions show four-fold rotational symmetry, evident from the four peaks in the cardinal directions.

\end{document}